\documentclass[12pt,a4paper]{article}
\usepackage{a4wide}
\usepackage[centertags]{amsmath}
\usepackage{amsfonts}
\usepackage{amssymb}
\usepackage{amsthm}
\numberwithin{equation}{section}
\newcommand{\hd}[1]{\phantom{ }^{\star_{#1}}}
\newcommand{\hde}{\phantom{ }^{\hat{\star}}}
\newcommand{\ttheta}{\tilde{\theta}}
\newcommand{\tphi}{\tilde{\varphi}}
\DeclareMathOperator{\tr}{Tr}
\allowdisplaybreaks[1]
\begin{document}
\pagestyle{empty}

\begin{flushright}
\small
DFTT 34/2001\\
IFT-UAM/CSIC-01-39\\
NORDITA-2001-89 HE\\
OSLO-TP 7-01\\
hep-th/0112126\\
\normalsize
\end{flushright}

\begin{center}
 

\vspace*{1.5cm}

{\Large {\bf GAUGE THEORIES FROM}}\\
\vskip 0.2cm 
{\Large {\bf WRAPPED AND FRACTIONAL BRANES \footnote{Work
partially supported by the European Commission
RTN programme HPRN-CT-2000-00131.}}}

\vspace{1.5cm}

{\bf P.~Di Vecchia ${}^{a}$, H.~Enger ${}^{b}$,
E.~Imeroni ${}^{c,\;a}$ and E.~Lozano--Tellechea ${}^{d,\;a}$}\\

\vskip 0.4cm
${}^{a}$\ \emph{NORDITA,
Blegdamsvej 17, DK-2100 Copenhagen \O, Denmark} \\
\vskip 0.4cm
${}^{b}$\ \emph{Department of Physics, University of Oslo,\\
P.O. Box 1048 Blindern, N-0316 Oslo, Norway}\\
\vskip 0.4cm
${}^{c}$\ \emph{Dipartimento di Fisica Teorica, Universit\`a di Torino\\
and I.N.F.N., Sezione di Torino, Via P. Giuria 1, I-10125 Torino, Italy}
\vskip 0.4cm
${}^{d}$\ \emph{Instituto de F\'{\i}sica Te\'orica, C-XVI,
Universidad Aut\'onoma de Madrid, \\
E-28049 Madrid, Spain}

\vspace{2cm}

{\bf Abstract}
\end{center}
\begin{quotation}
\small
We compare two applications of the gauge/gravity correspondence to a
non conformal gauge theory, based respectively on the study of
D-branes wrapped on supersymmetric cycles and of fractional D-branes
on orbifolds. We study two brane systems whose geometry is dual to
$\mathcal{N}=4\,$, $D=2+1$ super Yang--Mills theory, the first one
describing D4-branes wrapped on a two-sphere inside a Calabi--Yau
two-fold and the second one corresponding to a system of fractional
D2/D6-branes on the orbifold $\mathbb{R}^4/\mathbb{Z}_2\,$. By probing
both geometries we recover the exact perturbative running coupling
constant and metric on the moduli space of the gauge theory. We also
find a general expression for the running coupling constant of the
gauge theory in terms of the ``stringy volume'' of the two-cycle which
is involved in both types of brane systems.
\end{quotation}
\newpage
\pagestyle{plain}

\tableofcontents

\vspace{0.5cm}

\section{Introduction}

The gauge/gravity correspondence has its origin on the fact that, on
the one hand, D-branes are classical solutions of the low-energy
string effective action and, on the other hand, they have a gauge
theory living in their world-volume. This means that the low-energy
dynamics of D-branes can be used to determine the properties of the
gauge theory and vice-versa.

The most successful realization of this correspondence is the Maldacena
conjecture \cite{Maldacena:1997re}, confirmed by all subsequent
studies, according to which ten-dimensional type IIB string theory
compactified on $AdS_5\times S^5$ is dual to $\mathcal{N}=4$ super
Yang--Mills theory in four-dimensional Minkowski spacetime.

However, $\mathcal{N}=4$ super Yang--Mills in four dimensions is a rather
special theory due to its conformal properties and its high amount of
supersymmetry. Therefore, a lot of effort has been recently devoted to
find possible extensions of the Maldacena duality to non conformal and
less supersymmetric gauge theories, or at least to use the low-energy
brane dynamics to extract information on the properties of such more
realistic theories.

Two approaches to this problem have been largely pursued, based
respectively on the study of:
\begin{itemize}
\item D-branes \emph{wrapped} on supersymmetric cycles \cite{MN};
\item \emph{Fractional} D-branes on orbifolds 
\cite{Douglas:1996sw,Johnson:1996py,Diaconescu:1998br,Klebanov:1999rd}\footnote{Apart
from the two approaches that we consider in this paper,
other interesting ones are based on the study
of fractional D-branes on conifolds (see Ref.~\cite{Herzog:2001xk} and 
Ref.s therein)
and of M-branes wrapped on Riemann surfaces 
\cite{Fayyazuddin:1999zu,Fayyazuddin:2000em,Brinne:2000fh}.}. 
\end{itemize}
After ${\cal{N}}=4$ super Yang--Mills, the next natural system to
consider is $\mathcal{N}=2$ super Yang--Mills theory in four
dimensions, also with matter. This system has been studied, on the one 
hand, by considering fractional D3-branes on orbifolds
\cite{Bertolini:2001dk,Polchinski:2001mx,Billo:2001vg} and systems made of
fractional D3/D7-branes \cite{Grana:2001xn,Bertolini:2001qa}, and,
on the other hand,
by considering D5-branes wrapped on a two-cycle inside a Calabi--Yau
two-fold \cite{Gauntlett:2001ps,Bigazzi:2001aj}.
The low-energy dynamics of wrapped branes has also been recently used to
study other gauge theories
\cite{Maldacena:2000yy,Nunez:2001pt,Gomis:2001vk,Edelstein:2001pu,Gomis:2001vg,Gomis:2001aa,Gauntlett:2001ur}.

The use of fractional and wrapped branes presents some interesting
similarities that are not surprising since fractional branes on
orbifolds can be seen as D-branes wrapped on cycles that are
vanishing in the orbifold limit of the ALE space 
which corresponds to the blow-up of the orbifold space 
\cite{Douglas:1996sw,Johnson:1996py,Diaconescu:1998br}.
In fact, by probing the supergravity solutions describing
the two types of systems, one is able to recover all the relevant
\emph{perturbative} information on the Coulomb branch of the
gauge theory living on the branes, namely the running 
coupling constant and the metric on the moduli space of the
theory. 

These two approaches, however, have not been able to provide
information on the \emph{nonperturbative} features of the gauge
theories, as for instance on the instanton contribution to the moduli
space of $\mathcal{N}=2$ super Yang--Mills in four dimensions. This is
related to the existence at short distance of an
enhan{\c{c}}on~\cite{Johnson:2000qt} where the supergravity solution
becomes inconsistent, because at this distance the probe brane becomes
tensionless, signalling the appearance of new massless degrees of
freedom. This means that the supergravity approximation is not valid
anymore and that the region inside the enhan{\c{c}}on is excised. This
fact prevents to get information on the strong coupling regime of the
gauge theory living in the world-volume of the branes, that is in fact
determined by what happens inside the enhan{\c{c}}on. To overcome this
problem one must presumably also include the new massless degrees of
freedom, as attempted for instance in Ref.~\cite{Wijnholt:2001us}\footnote{For recent
developments concerning the physics of the enhan\c con see for
instance
Ref.s~\cite{Petrini:2001fk,Johnson:2001wm,Merlatti:2001gd}.}.

In this paper we will apply both approaches discussed above
to the study of $\mathcal{N}=4\,$, $D=2+1$ super Yang--Mills theory,
which is a theory with 8 Poincar\'e supercharges.
The interest in this theory resides mainly in the fact that
its properties, perturbative and nonperturbative, are well known
\cite{Alvarez-Gaume:1981hm,Seiberg:1996nz}. This
is also the theory where the enhan{\c{c}}on was first
found~\cite{Johnson:2000qt} using a different approach based on the
study of D6-branes wrapped on $K3$ surfaces.

We will first consider a system made up of $N$ D4-branes wrapped on
a two-cycle inside a Calabi--Yau two-fold. The crucial property
of this system, as of any other system of branes wrapped on 
supersymmetric cycles,
is that the geometrical structure of the background
forces the gauge theory living on the world-volume of the branes
to be partially \emph{topologically twisted} \cite{BVS} and
this allows to preserve the desired amount of supersymmetry.
To find the supergravity solution describing the D4-branes,
we will use the techniques introduced in Ref.~\cite{MN},
which amount to find a solution of a lower dimensional
gauged supergravity and then uplift it to ten or eleven dimensions.
We will then use the uplifted solution in a probe computation in order
to extract information on the Coulomb branch of the gauge theory which 
lives on the flat
three-dimensional part of the world-volume of the brane, which
is pure $\mathcal{N}=4\,$, $D=2+1$ super Yang--Mills with gauge group 
$SU(N)\,$.

Then, we will consider a system made of $N$ fractional D2-branes and
$M$ D6-branes on the orbifold $\mathbb{R}^4/\mathbb{Z}_2\,$ and,
solving explicitly the equations of motion of type IIA supergravity,
we will find the corresponding classical solution. The probe
computation will give us information on the same three-dimensional
theory, now also coupled to $M$ hypermultiplets in the fundamental
representation of the gauge group.

In both approaches we find that, as in other cases, the probe analysis
correctly reproduces the perturbative part of the moduli space, giving
the exact running coupling constant of ${\cal{N}}=4$ super Yang--Mills in
three dimensions, but is unable to give the instanton contribution.
This analysis allows us to make some comments on the relation between
the two solutions and to see that in both cases the gauge coupling
constant can be obtained from a common expression representing the
``stringy volume'' of the two-cycle on which the branes are
wrapped. Moreover, in both cases the locus where the ``stringy
volume'' vanishes corresponds to the point where the Calabi--Yau
two-fold in which the cycle is embedded manifests an enhanced gauge
symmetry, which is the origin of the enhan\c con mechanism.

The structure of this paper is as follows. Section \ref{wrappedD4} and
\ref{fractionalsystem} are organized in an entirely parallel way,
and can be read independently from each other. They describe the two 
different brane systems that we study in order to get information
on $\mathcal{N}=4\,$, $D=2+1$ super Yang--Mills theory from
supergravity, namely a system of $N$ D4-branes wrapped on $S^2$
(in section \ref{wrappedD4}) and a system of $N$ fractional D2-branes
and $M$ D6-branes on the orbifold $\mathbb{R}^4/\mathbb{Z}_2$
(in section \ref{fractionalsystem}). In section \ref{discussion}, we 
discuss
and comment the results of the previous two sections.
Many details of the various computations are given in the appendices.
In appendix \ref{appsugrasol} we fix the conventions and
discuss in detail how the two supergravity solutions were found.
In appendix \ref{wvact} we discuss the world-volume
actions for fractional branes. Finally, in appendix \ref{gaugerun}
we give some details about the perturbative computation of
the running coupling constant of the gauge
theory that we consider.


\section{D4-branes wrapped on $S^2$}\label{wrappedD4}

\subsection{Setup}\label{wsetup}

In this section we are going to consider a system made of $N$
D4-branes with two longitudinal directions wrapped on
a two-sphere.

As discussed in Ref.~\cite{BVS}, the gauge theory living on the
world-volume of wrapped branes has to be topologically twisted.  In
this subsection we want to determine the topological twist that is
needed in order to obtain at low-energy on the flat part of the
world-volume of the D4-branes $\mathcal{N}=4$ super Yang--Mills theory
in three space-time dimensions, that is a theory with 8
supercharges. The twist which preserves 8 supercharges is exactly the
one imposed by the geometrical structure of the background when the
two-sphere is seen as a nontrivial two-cycle inside a Calabi--Yau two-fold.

The configuration that we are going to study
is schematically shown in the following table, where
the symbols --\,, $\smallfrown$ and $\cdot$ represent respectively
unwrapped world-volume directions, wrapped world-volume directions and transverse directions:
\begin{center}
\begin{tabular}{|c|c|c|c|c|c|c|c|c|c|c|}
\multicolumn{4}{c}{ }
&\multicolumn{4}{c}{$\overbrace{\phantom{\qquad\qquad\quad}}^{CY}$}
&\multicolumn{3}{c}{ }\\
\hline
&\multicolumn{3}{|c|}{$\mathbb{R}^{1,2}$}&\multicolumn{2}{|c|}{$S^2$}&
\multicolumn{2}{|c|}{$N_2$}&\multicolumn{3}{|c|}{$\mathbb{R}^{3}$}\\
\hline
D4 &--&--&--&$\smallfrown$&$\smallfrown$&$\cdot$&$\cdot$&$\cdot$&$\cdot$&$\cdot$\\
\hline
\end{tabular}
\end{center}

In flat space, the presence of the D4-brane breaks spacetime Lorentz 
invariance in the following way: 
$SO(1,9)\longrightarrow SO(1,4)\times SO(5)_R\,$. The fact that the
D4-brane is wrapped on $S^2$ introduces an additional breaking of
$SO(1,4)$ into $SO(1,2) \times SO(2)_{S^2}$. The twist is then
introduced by breaking the $R$-symmetry group $SO(5)_R$ into 
$ SO(2)_G \times SO(3) $ and by identifying $SO(2)_G$ with $SO(2)_{S^2}$. 
In conclusion our configuration breaks the original $SO(1,4) \times
SO(5)_R$ into
$SO(1,2) \times SO(2)_{S^2} \times SO(2)_G \times SO(3)\,$ with the
two $SO(2)$ groups identified.
The fields of the gauge theory living on the wrapped D4-branes transform 
according to the following representations of the above groups:
\begin{center}
\begin{tabular}{|c|ccc|ccc|}
\hline
&$SO(1,4)$&$\longrightarrow$&$SO(1,2) \times SO(2)_{S^2}$
&$SO(5)_R$&$\longrightarrow$&$SO(2)_G \times SO(3)$\\
\hline
Vector&$\mathbf{5}$&$\longrightarrow$&$(\mathbf{3},\mathbf{1}) \oplus (\mathbf{1},\mathbf{2})$
&$\mathbf{1}$&$\longrightarrow$&$(\mathbf{1},\mathbf{1})$\\
\hline
Scalars&$\mathbf{1}$&$\longrightarrow$&$(\mathbf{1},\mathbf{1})$
&$\mathbf{5}$&$\longrightarrow$&$(\mathbf{1},\mathbf{3}) \oplus (\mathbf{2},\mathbf{1})$\\
\hline
Fermions&$\mathbf{4}$&$\longrightarrow$&$(\mathbf{2},\mathbf{+}) \oplus (\mathbf{2},\mathbf{-})$
&$\mathbf{4}$&$\longrightarrow$&$(\mathbf{+},\mathbf{2}) \oplus (\mathbf{-},\mathbf{2})$\\
\hline
\end{tabular}
\end{center}

Since we are interested in the
three-dimensional theory living on the \emph{flat} part of the world-volume
at very low energies,
we must keep only the massless states, which are 
the ones transforming as singlets under
$SO(2)_D\equiv(SO(2)_{S^2}\times SO(2)_G)_{\text{diag}}\,$:
\begin{center}
\begin{tabular}{|c|c|}
\hline
&$SO(1,2) \times SO(2)_D \times SO(3)$\\
\hline
Vector&$(\mathbf{3},\mathbf{1},\mathbf{1})$\\
\hline
Scalars&$(\mathbf{1},\mathbf{1},\mathbf{3})$\\
\hline
Fermions&$2\times(\mathbf{2},\mathbf{1},\mathbf{2})$\\
\hline
\end{tabular}
\end{center}
These states form exactly the vector multiplet
of $\mathcal{N}=4\,$, $D=2+1$ super Yang--Mills theory.


\subsection{The supergravity solution}{\label{SectTSG-D4onS2}}

In this subsection we will construct a supergravity solution
describing the system just introduced,
made of $N$ D4-branes with two world-volume directions wrapped on $S^2\,$.
One could in principle work in ten-dimensional type IIA supergravity, write
a suitable Ansatz for such a system,
then solve the equations of motion and find the solution. This
is, however, not a simple task because it is not easy to implement
directly in ten dimensions the topological twist that we have discussed in the
previous subsection from the point of view of the gauge theory living 
on the brane. One has to proceed in a longer, but
more straightforward way that has been introduced in Ref.~\cite{MN}. 
Instead of working directly in the ten dimensional theory one starts by 
considering, for the case of a $p$-brane, a $(p+2)$-dimensional gauged supergravity
theory that is obtained by compactifying the original $D$-dimensional
theory (where of course $D=10$ for the case of a D-brane or NS5-brane
and $D=11$ for the case of an M-brane)
on $S^{D-p-2}$. The isometry group \mbox{$SO(D-p-1)$} of $S^{D-p-2}$
corresponds to the $R$-symmetry group that we discussed in the previous
subsection. In gauged supergravity the
$R$-symmetry group is gauged so that the theory contains $SO(D-p-1)$ gauge
fields. In this theory one looks for a domain wall solution that
preserves the desired amount of supersymmetry and breaks the original 
$R$-symmetry group in a way that implements the correct twist.
In fact, in
gauged supergravity the supersymmetry preserving condition contains
also the gauge fields and can schematically be written as 
$(\partial_\mu+\omega_\mu+A_\mu)\ \epsilon =0$. 
The discussed twist corresponds to the identification of
some of the gauge fields with the spin
connection of the manifold around which the brane is wrapped,
$A_\mu=-\omega_\mu\,$, so that the request of finding covariantly constant 
spinors is equivalent to that of just finding constant spinors.
Once the solution with the correct
properties has been found, the last step is to uplift it to $D$
dimensions by using the formulas given in
Ref.s~\cite{Cv1,Cv2}.

In the following, in order to avoid many new calculations, we do not 
use directly a 6-dimensional gauged supergravity
as it would be natural for a D4-brane. We will instead
proceed in slightly different way by exploiting the fact that the
solution of seven-dimensional gauged supergravity corresponding
in eleven dimensions to an 
M5-brane wrapped on $S^2\,$ and preserving $8$ supercharges has already
been constructed~\cite{MN}. Therefore we proceed as follows.
We start from the solution of 7-dimensional gauged supergravity
given in Ref.~\cite{MN}, and uplifting it to eleven dimensions using the
formulas found in Ref.~\cite{Cv1} we obtain a solution of
11-dimensional supergravity
describing $N$ M5-branes wrapped on $S^2$. Finally, upon compactification
to ten dimensions we get the desired solution describing
$N$ D4-branes wrapped on $S^2$. The details of this procedure are
given explicitly in Appendix~\ref{D4onS2SugraSolution}. Here we write
directly the ten-dimensional solution in the string frame, which reads\footnote{This
solution was partially
given in appendix~7.4 of Ref.~\cite{MN}.}:
\begin{subequations}
\label{10dimsolution}
\begin{align}
  ds_{\text{st}}^2&=\left(\frac{R_A}{R_0}\right)^3\Delta^{1/2}e^{3\rho}\, 
          \eta_{\alpha\beta}d\xi^\alpha d\xi^\beta+
          \frac{R_A^3}{R_0}\Delta^{1/2}e^{\rho}
          (e^{2\rho}-{\textstyle\frac{1}{4}})
          (d\ttheta^2+\sin^2\ttheta d\tphi^2)  \nonumber \\ 
        &+\frac{R_A^3}{4R_0}\Delta^{-1/2}e^\rho\left(
          \frac{4\Delta}{e^{5\lambda}}d\rho^2+\Delta d\chi^2+
          \cos^2\chi(d\theta^2+\sin^2\theta d\varphi^2) +   
          e^{5\lambda}\sin^2\chi\left(d\psi+\cos\ttheta d\tphi\right)^2  
          \right)\, , \\ 
  e^{2\phi}&=\left(\frac{R_A}{R_0}\right)^3\Delta^{1/2}e^{3\rho}\, ,  \\ 
  C_3&=\frac{R_A^3}{8}\, \frac{e^{5\lambda}\cos^3\chi\cos\theta\sin\ttheta}
        {\Delta}\, d\ttheta\wedge d\tphi\wedge d\varphi  \nonumber \\ 
      &+\frac{R_A^3}{8}\, \frac{e^{5\lambda}
        (\Delta+2)\cos^2\chi\sin\chi\cos\theta}{\Delta^2}\, 
        d\chi\wedge d\varphi\wedge\left(d\psi+\cos\ttheta d\tphi\right) \nonumber \\
      &+\frac{R_A^3}{8}\, \frac{\partial_\rho(e^{5\lambda})
        \cos^3\chi\sin^2\chi\cos\theta}{\Delta^2}\, 
        d\rho\wedge d\varphi\wedge\left(d\psi+\cos\ttheta d\tphi\right)\, ,  
\end{align}
\end{subequations}
where the functions $e^{5\lambda}$ and $\Delta$ entering the solution 
are given by:
\begin{subequations}
\label{e5lambdaDelta}
\begin{align}
  e^{5\lambda}&=\frac{e^{2\rho}+ke^{-2\rho}-\frac{1}{2}}
                {e^{2\rho}-\frac{1}{4}}\, , \\
  \Delta&=e^{5\lambda}\cos^2\chi+\sin^2\chi\, . 
\end{align}
\end{subequations}
Before proceeding let us give a short explanation of the various 
coordinates and constants appearing in
eq.s~\eqref{10dimsolution}-\eqref{e5lambdaDelta}:
\begin{itemize}
\item $\xi^{\alpha,\beta}$ ($\alpha,\beta=0,1,2$) are the coordinates
along the unwrapped brane world-volume;
\item $\ttheta\,,\tphi$ are the coordinates along the wrapped world-volume;
\item $\rho$ is a radial coordinate transverse to the brane;
\item $\chi,\theta,\varphi,\psi$ parameterize the ``twisted'' four-sphere
transverse to the brane;
\item $R_A$ is the radius of the $AdS_7$ space appearing in the near
horizon geometry of the usual ``flat'' M5-brane solution
(see appendix \ref{D4onS2SugraSolution}),
which is given in terms of ten dimensional quantities by
$R_A=2\sqrt{\alpha'}(\pi g_s N)^{1/3}\,$;
\item $R_0$ is an arbitrary integration constant with dimension of a 
length that we will show to set the scale of the
radius of the $S^2$ on which the D4-branes are
wrapped;
\item $k$ is a dimensionless integration constant.
\end{itemize}
All coordinates are dimensionless except $\xi^\alpha$ which
have dimension of a length. 

A D4-brane is coupled naturally to a $5$-form potential while the solution
given above contains a RR $3$-form potential. However, the latter is
related to $C_5$ by the duality relation $dC_5=\hd{} dC_3$ (in the
string frame). By using it we get:
\begin{equation}
\begin{split}
  C_5=&\ \frac{R_A^6}{R_0^4}\, \Delta e^{4\rho}
       \left(e^{2\rho}-\tfrac{1}{4}\right) \sin\ttheta\, 
       d\xi^0\wedge d\xi^1\wedge d\xi^2\wedge d\ttheta\wedge d\tphi  \\ 
     &-\frac{R_A^6}{R_0^4}\, \frac{1}{2}e^{4\rho}\sin^2\chi 
        \ d\xi^0\wedge d\xi^1\wedge d\xi^2\wedge d\rho
        \wedge\left(d\psi+\cos\ttheta d\tphi\right)  \\
     &-\frac{R_A^6}{R_0^4}\, \frac{1}{8}e^{4\rho}e^{5\lambda}\sin(2\chi)
        \ d\xi^0\wedge d\xi^1\wedge d\xi^2\wedge d\chi
        \wedge\left(d\psi+\cos\ttheta d\tphi\right)\, .
\end{split}
\end{equation}


\subsubsection*{A change of coordinates}

The supergravity solution for the D4-branes wrapped on $S^2$ as
given in eq.~\eqref{10dimsolution} is written in a way in which 
the role of the different coordinates and factors is not immediately
clear. The first thing that we can do in order to clarify the role
of the various terms appearing in the solution is to extract the warp
factors for the longitudinal and transverse part of the metric in the
string frame. They are given in terms of a function $H$ that for a
D4-brane is related to the dilaton through the following relation:
\begin{equation}
\label{myH}
   H=e^{-4\phi}=\left(\frac{R_0}{R_A}\right)^6\Delta^{-1}e^{-6\rho}\,.
\end{equation}
Using the previous definition of $H$, one can immediately see that the 
dependence on $H$ of the four longitudinal unwrapped directions of the
metric is the one corresponding to four ``flat'' world-volume
directions:
$H^{-1/2}\eta_{\alpha\beta}d\xi^\alpha d\xi^\beta\,$, as expected.
We also expect three transverse
directions ($\theta\,$, $\varphi$ and
a suitable combination of $\rho$ and $\chi\,$) to be
flat, apart from the usual warp factor $H^{1/2}\,$.
This can be seen to be correct\footnote{See
also Ref.s~\cite{Gauntlett:2001qs,Babington:2001nh} for related 
discussions.}
by using instead of the coordinates $\rho$
and $\chi$ the following new coordinates:
\begin{equation}\label{mycoord}
\begin{cases}
   r=\frac{R_A^3}{2R_0^2}e^{2\rho}\cos \chi \\ \\
   \sigma=\frac{R_A^3}{2R_0^2}\left[e^{2\rho}\left(e^{2\rho}-\frac{1}{4}\right)
         e^{5\lambda}\right]^{1/2}\sin \chi
\end{cases}
\end{equation}
which have dimensions of a length.
In terms of the new coordinates in eq.~\eqref{mycoord},
the solution for the metric, dilaton and
R-R 5-form becomes:
\begin{subequations}\label{newsol}
\begin{align}
   ds_{\text{st}}^2&=H^{-1/2}\left[\eta_{\alpha\beta}d\xi^\alpha d\xi^\beta+\mathcal{Z}R_0^2
        (d\ttheta^2+\sin^2\ttheta d\tphi^2)\right]\nonumber\\
        &\qquad\quad+H^{1/2}\left[dr^2+r^2\left(d\theta^2+\sin^2\theta d\varphi^2\right)
        +\frac{1}{\mathcal{Z}}\left(d\sigma^2+\sigma^2\left(d\psi+\cos\ttheta d\tphi\right)^2\right)\right]\,,
        \label{newsolm}\\
   e^\phi&=H^{-1/4}\,,\label{newsold}\\
   C_5&=d\xi^0\wedge d\xi^1\wedge d\xi^2\wedge
       \left[\frac{1}{H}\mathcal{Z}R_0^2\sin\ttheta\ d\ttheta\wedge d\tphi
       -\frac{1}{\mathcal{Z}}\sigma d\sigma\wedge\left(d\psi+\cos\ttheta d\tphi\right)\right]\,,\label{newsolC5}
\end{align}
\end{subequations}
where the functions $H$ and $\mathcal{Z}$ are (implicitly) defined as:
\begin{equation}
   H(r,\sigma)=\left(\frac{R_0}{R_A}\right)^6\Delta^{-1}(r,\sigma)e^{-6\rho(r,\sigma)}\,,\qquad
   \mathcal{Z}(r,\sigma)=e^{-2\rho(r,\sigma)}\left(e^{2\rho(r,\sigma)}-\frac{1}{4}\right)\,.
\end{equation}
In the form given in eq.~\eqref{newsol} the structure of the solution
is much clearer. First of all one can clearly distinguish the trivial
``flat'' part of the solution from the nontrivial part coming from the
internal directions of the four-dimensional Calabi--Yau space.  In this
sense, the coordinates $r$ and $\sigma$ that we have introduced
represent two radial directions, respectively in the ``flat''
transverse space and in the space $N_2$ transverse to the brane but
nontrivially fibered on the two-cycle on which the brane is
wrapped. Moreover, the function $\mathcal{Z}$ represents the ``running
volume'' of the two-cycle, with the constant $R_0$ being the radius of
the $S^2$ when $\mathcal{Z}=1\,$, while in the part of the metric
containing $\sigma$ and $\psi$ we can easily see the twist which, as
we have seen in section \ref{wsetup}, is required for having a
supersymmetric gauge theory living on the brane. Finally, also the R-R
potential has a quite standard part ($H^{-1}$ times the volume form of
the longitudinal space), plus an additional part due to the
twist.

Another change of coordinates can be implemented to extract some additional piece
of information about the solution. If we define a new coordinate $z$ and
a function $\widetilde{\mathcal{Z}}$ as follows:
\begin{equation}\label{ALEcoords}
\begin{cases}
   z = R_0 \left(1+\frac{\sigma^2}{R_0^2}\right)^{1/4}\\ \\
   \widetilde{\mathcal{Z}} = \mathcal{Z} \left(1+\frac{\sigma^2}{R_0^2}\right)^{-1/2}
\end{cases}
\end{equation}
the metric in eq.~\eqref{newsolm} becomes:
\begin{multline}\label{ALE}
   ds_{\text{st}}^2=H^{-1/2}\left\{\eta_{\alpha\beta}d\xi^\alpha 
        d\xi^\beta+\widetilde{\mathcal{Z}}\ z^2(d\ttheta^2+\sin^2\ttheta d\tphi^2)\right\}
        +H^{1/2}\Bigg\{dr^2+r^2\left(d\theta^2+\sin^2\theta d\varphi^2\right)\\
        +\frac{1}{\widetilde{\mathcal{Z}}}\left[4\left(1-\frac{R_0^4}{z^4}\right)^{-1}dz^2
        +z^2\left(1-\frac{R_0^4}{z^4}\right)
        \left(d\psi+\cos\ttheta d\tphi\right)^2\right]\Bigg\}\,.
\end{multline}
The metric we have obtained on the four-dimensional space spanned by
the coordinates $\{\ttheta,\tphi,z,\psi\}$ is that of a ``warped''
Eguchi--Hanson space \cite{Eguchi:1978gw}. This fact provides
additional evidence of the geometrical structure of the background:
the D4-branes are wrapped on the two-sphere, of radius $R_0\,$, inside
the simplest ALE space (which corresponds to the blow-up of an
$\mathbb{R}^4/\mathbb{Z}_2$ orbifold)\footnote{As an aside, notice
that, using eq.s~\eqref{myH}-\eqref{mycoord} and
eq.~\eqref{ALEcoords}, also the M5-brane solution
\eqref{11dimsolution} (from which we derived the D4-brane solution)
can be brought into a form analogous to the one given in
eq.~\eqref{newsol} or in eq.~\eqref{ALE}. These forms for the
classical solutions describing wrapped branes seem indeed to be quite
general. For instance, changes of coordinates similar to the ones
implemented here can be used to put in these forms also the solution
found in Ref.~\cite{Gauntlett:2001ps} for D5-branes wrapped on
$S^2\,$.}.

\subsection{Probing the wrapped brane solution}\label{wprobesect}

In order to extract information on the gauge theory living
on the D4-branes, we will study the dynamics of a probe
D4-brane wrapped on $S^2$ in the geometry generated by the solution
found in the previous subsection (see Ref.~\cite{Johnson:2000ch} for a review
on the probe technique). This will allow us to study
the Coulomb branch of pure $\mathcal{N}=4$ SYM theory in $2+1$
dimensions with gauge group $SU(N+1)\longrightarrow SU(N)\times U(1)$.
The world-volume action for a single D4-brane in the string frame
in the static gauge is
given by:
\begin{equation}\label{wprobe}
   S_{\text{probe}}=-\frac{T_{4}}{\kappa}
         \int{d^{3}\xi d\ttheta d\tphi e^{-\phi}
         \sqrt{-\det \left[G_{ab}+
         2\pi\alpha'F_{ab}\right]}}
      +\frac{T_{4}}{\kappa}\int_{{\cal M}_5}{\big(
         C_5+2\pi\alpha' C_3\wedge F\big)}\,, 
\end{equation}
where $a,b=\{0,1,2,\ttheta,\tphi\}$ and all the bulk fields are understood to be pullbacks onto the
brane world-volume.

Let us first compute the static potential
between the probe and the stack of $N$ D4-branes, simply by substituting the
solution \eqref{newsol} into eq.~\eqref{wprobe}.
The contribution of the Dirac--Born--Infeld part is given by: 
\begin{equation}
  e^{-\phi}\sqrt{-\det G_{ab}}=\sin\ttheta\frac{\mathcal{Z}R_0^2}{H}
     \left(1+\frac{\sigma^2H}{\mathcal{Z}^2R_0^2}\right)^{1/2}\,.
\end{equation}
Adding to it the Wess--Zumino part, whose contribution
is computed using the expression
\eqref{newsolC5} of the R-R 5-form, we get
the following expression for the static potential:
\begin{equation}\label{potential}
   S_{\text{pot}}=-\frac{T_{4}}{\kappa}
      \int d^{3}\xi d\ttheta d\tphi \sin\ttheta
      \frac{\mathcal{Z}R_0^2}{H}
      \left[\left(1+\frac{\sigma^2H}{\mathcal{Z}^2R_0^2}\right)^{1/2}-1\right]\,.
\end{equation}
We see that in general there is a force between
the branes, and this means that the configuration is not
supersymmetric. This had to be expected in some way because we are allowing
the probe brane to move in \emph{all} its transverse
directions, including also the ones which are inside the Calabi--Yau space.
If instead we allow the probe brane to move only in the
``flat'' part of the transverse space spanned by $\{r,\theta,\varphi\}\,$,
keeping it fixed at the locus $\sigma=0$ in the ``internal'' transverse space,
we see that the potential \eqref{potential} vanishes,
yielding a supersymmetric configuration.
Therefore in the following we will always work at
the ``supersymmetric locus'' $\sigma=0$.

In order to study the dynamics of the probe brane, we will allow the
transverse coordinates $X^i=\{r,\theta,\phi\}$ to depend on the
``flat'' world-volume coordinates $\xi^\alpha$ but not on the 
``wrapped'' ones $x$ and $y$. Moreover, the gauge field $F_{\alpha\beta}$
is defined to be nonvanishing only on the ``flat'' part of the world-volume.
Let us start from the DBI part of the action in eq.~\eqref{wprobe}. By 
expanding the determinant, we find:
\begin{multline}
   S_{\text{DBI}}\simeq-\frac{T_{4}}{\kappa}\int d^{3}\xi d\ttheta d\tphi
      e^{-\phi}\sqrt{-\det G_{ab}}\\
      \times\left\{1+\frac{1}{2}G^{\alpha\beta}G_{ij}\partial_\alpha X^i\partial_\beta X^j
      +\frac{(2\pi\alpha')^2}{4}G^{\alpha\gamma}G^{\beta\delta}
      F_{\alpha\beta}F_{\gamma\delta}\right\}\,.
\end{multline}
Inserting the expressions \eqref{newsolm} and \eqref{newsold}
for the metric and dilaton we get:
\begin{equation}
   S_{\text{DBI}}=-\frac{T_{4}}{2\kappa}\int d^{3}\xi d\ttheta d\tphi \sin\ttheta
      \frac{\mathcal{Z}R_0^2}{H}
      \left\{1+\frac{1}{2}H 
      \big[(\partial r)^2+r^2\left((\partial\theta)^2+\sin^2\theta(\partial\varphi)^2\right)\big]
      +\frac{(2\pi\alpha')^2}{4}H F^2\right\}\,,
\end{equation}
where we have included an additional factor of $1/2$ due to the normalization
of the generators of the gauge group. 

Notice that when $\mathcal{Z}=0\,$, the effective tension
of the brane vanishes, meaning that an enhan\c con mechanism
is taking place \cite{Johnson:2000qt}. Since, in order to preserve
supersymmetry, we have fixed $\sigma=0$, the enhan{\c{c}}on radius is
given by:
\begin{equation}
   r_{\text{e}}=\frac{R_A^3}{8R_0^2}=\frac{\pi g_s (\alpha')^{3/2}N}{R_0^2}\,.
\end{equation}
In fact, this seems to be a general feature (albeit somewhat unnoticed
in the literature) of the supergravity solutions corresponding to
D-branes wrapped on cycles\footnote{Notice however that the nature and
location of the singularities of the metric depend on the value of the
constant $k$ appearing in eq.~\eqref{e5lambdaDelta}, as discussed in
very similar cases in
Ref.s~\cite{MN,Gauntlett:2001ps,Gauntlett:2001ur}. Nonetheless, gauge
theory physics as seen by the brane probe at the supersymmetric locus
is independent of $k\,$, and only feels the existence of the enhan\c
con.}. The fact that the solution is no longer valid inside the
enhan\c con radius seems to prevent us from getting nonperturbative
information on the gauge theory under study.

The transverse scalars have to be interpreted as Higgs fields
for the gauge theory living on the brane: $X^i=2\pi\alpha'\Phi^i$.
Then, defining $\mu$ such that $r=2\pi\alpha'\mu$
and integrating over the volume of the two-sphere on which the brane is
wrapped, we obtain the final expression for the DBI part:
\begin{multline}\label{wdbi}
   S_{\text{DBI}}=-\frac{4\pi T_{4}}{2\kappa}
      \int d^{3}\xi 
      \frac{\mathcal{Z}R_0^2}{H}\\
      \times\left\{1+\frac{(2\pi\alpha')^2}{2}H 
      \big[(\partial\mu)^2+\mu^2\left((\partial\theta)^2+\sin^2\theta(\partial\varphi)^2\right)\big]
      +\frac{(2\pi\alpha')^2}{4}H F^2\right\}\,.
\end{multline}
Turning now to the WZ part, the pullback of $C_3$ is given by:
\begin{equation}
   C_3=\frac{1}{8}R_A^3\cos\theta\sin\ttheta\partial_\alpha \varphi d\xi^\alpha\wedge d\ttheta\wedge d\tphi \,. 
\end{equation}
Then from eq.~\eqref{wprobe} we get:
\begin{equation}\label{wwz}
\begin{split}
   S_{\text{WZ}}&=\frac{T_4}{\kappa}\int d^3\xi d\ttheta d\tphi \sin\ttheta\left\{
      \frac{\mathcal{Z}R_0^2}{H}
      +\frac{2\pi\alpha' R_A^3}{16}\cos\theta\varepsilon^{\alpha\beta\gamma}
      \partial_\alpha \varphi F_{\beta\gamma}\right\}\\
      &=\frac{4\pi T_4}{\kappa}\int d^3\xi \left\{
      \frac{\mathcal{Z}R_0^2}{H}
      +\frac{2\pi\alpha' R_A^3}{16}\cos\theta\varepsilon^{\alpha\beta\gamma}
      \partial_\alpha \varphi F_{\beta\gamma}\right\}\\
\end{split}
\end{equation}
Putting eq.s~\eqref{wdbi} and \eqref{wwz} together and
substituting the expressions for $T_4\,$, $\kappa\,$,
$R_A$ and for the function $\mathcal{Z}$, the probe action finally
becomes:
\begin{multline}
\label{wfinprobe}
   S_{\text{probe}}=-\frac{R_0^2}{2\pi g_s\sqrt{\alpha'}}
      \int d^{3}\xi \left(1-\frac{g_s\sqrt{\alpha'}N}{2R_0^2\mu}\right)
      \Bigg\{\frac{1}{2} 
      \big[(\partial\mu)^2+\mu^2\left((\partial\theta)^2+\sin^2\theta(\partial\varphi)^2\right)\big]
      +\frac{1}{4} F^2\Bigg\}\\
      +\frac{N}{8\pi}\int d^3\xi \cos\theta
      \varepsilon^{\alpha\beta\gamma}\partial_\alpha\varphi F_{\beta\gamma}\,.
\end{multline}

From the coefficient of $F^2\,$ in eq.~\eqref{wfinprobe} we can
read the running gauge coupling constant of the three-dimensional
gauge theory as a function of the scale $\mu\,$. Defining the bare
coupling as:
\begin{equation}\label{wbare}
   g^2_{\text{YM}}=\frac{2\pi g_s \sqrt{\alpha'}}{R_0^2}\,,
\end{equation}
the running coupling constant is given by:
\begin{equation}\label{wrunning}
   \frac{1}{g^2_{\text{YM}}(\mu)}=
   \frac{1}{g^2_{\text{YM}}}\left(1-\frac{g^2_{\text{YM}}N}{4\pi\mu}\right)\,, 
\end{equation}
in perfect agreement with gauge theory expectations, as shown in
appendix \ref{gaugerun}. 

Eq.~\eqref{wfinprobe} does not give explicitly the full metric
on the moduli space of $\mathcal{N}=4\,$, $D=2+1$ SYM theory. In fact
such a metric must be hyperK\"ahler \cite{Alvarez-Gaume:1981hm} and in
eq.~(\ref{wfinprobe}) we have only three moduli and not four as it
should be in a hyperK\"ahler metric. We need an extra modulus that
can be obtained by dualising the vector field. In order to do that,
we regard the original action in eq.~\eqref{wfinprobe}
as a function of $F_{\alpha\beta}$ and we add to it a term:
\begin{equation}\label{addterm}
   - \int\ \Sigma\ dF\,,
\end{equation}
so that the equation of motion for the auxiliary field $\Sigma$
enforces the Bianchi identity for $F$ on shell. By partially integrating the
additional term in eq.~\eqref{addterm}, we are left with the 
following action:
\begin{multline}\label{dual1}
    S_{\text{probe}}=-
     \int d^3 \xi\ \frac{1}{g^2_{\text{YM}}(\mu)}
     \Bigg\{\frac{1}{2} 
      \big[(\partial\mu)^2+\mu^2\left((\partial\theta)^2+\sin^2\theta(\partial\varphi)^2\right)\big]
      +\frac{1}{4} F^2\Bigg\} \\
     +\frac{N}{8\pi}\int d^3 \xi \cos\theta 
     \varepsilon^{\alpha\beta\gamma}\partial_\alpha\varphi F_{\beta\gamma}
     +\frac{1}{2}\int d^3 \xi 
     \varepsilon^{\alpha\beta\gamma}\partial_\alpha\Sigma F_{\beta\gamma}\,.
\end{multline} 
We can then eliminate $F$ by means of its equation of motion that
follows from eq.~\eqref{dual1}:
\begin{equation}
   F_{\beta\gamma}=g^2_{\text{YM}}(\mu)\varepsilon^{\alpha\beta\gamma}
      \left[\frac{N}{4\pi}\cos\theta\partial_\alpha\varphi
      +\partial_\alpha\Sigma\right]\,,
\end{equation}
and we arrive at an action that contains four moduli, given by:
\begin{multline}
    S_{\text{probe}}= -\frac{1}{2}
      \int d^3 \xi \Bigg\{
      \frac{1}{g^2_{\text{YM}}(\mu)} 
      \big[(\partial\mu)^2+\mu^2\left((\partial\theta)^2+\sin^2\theta(\partial\varphi)^2\right)\big]\\
      +g^2_{\text{YM}}(\mu)\left(\frac{N\cos\theta}{4\pi}\partial\varphi
      +\partial\Sigma\right)^2\Bigg\}\,.
\end{multline} 
The complete metric on the moduli space $\mathcal{M}$ of
the gauge theory, in terms of the 4 scalars $\mu\,$, $\theta\,$,
$\phi\,$ and $\Sigma$ is finally given by:
\begin{equation}\label{wMetricMS}
   ds^2_{\mathcal{M}}=
     \frac{1}{g^2_{\text{YM}}(\mu)}
     \left(d\mu^2+\mu^2d\Omega^2\right)
     +g^2_{\text{YM}}(\mu)
     \left(d\Sigma+
     \frac{N\cos\theta}{4\pi}d\varphi\right)^2\,,
\end{equation}
where $d\Omega^2=d\theta^2+\sin^2\theta d\varphi^2\,$. The metric in eq.
\eqref{wMetricMS} is indeed hyperK\"ahler since it has
precisely the form of the Taub-NUT metric
\cite{Hawking:1977jb}. However, because of the form given in
eq.~\eqref{wrunning} of the function $g_{\text{YM}}(\mu)$
our metric has a ``negative mass'' and thus is singular. This is due
to the fact that in our probe analysis we are only able to reproduce
the perturbative behaviour of the gauge theory. As
discussed in Ref.s~\cite{Johnson:2000qt,Johnson:2000ch}, the complete 
metric should also include the instanton contribution, becoming
a completely nonsingular generalization of the Atiyah-Hitchin
metric.


\section{Fractional D2/D6-brane system}\label{fractionalsystem}

\subsection{Setup}\label{setup}

In this section we consider a system of fractional branes
on the orbifold:
\begin{equation}\label{orb}
  \mathbb{R}^{1,5}\times\mathbb{R}^4/\mathbb{Z}_2\, , 
\end{equation}
where $\mathbb{Z}_2$ acts by changing sign to the last four coordinates:
\begin{equation*}
  \left\{x^6, x^7, x^8, x^9\right\}
  \longrightarrow\left\{-x^6,-x^7,-x^8,-x^9\right\} \, .
\end{equation*}

To be precise, we are going to study a configuration of type IIA string
theory\footnote{Classical solutions describing fractional
D-branes in type IIA orbifolds were constructed in Ref.~\cite{Frau:2001gk}.},
made of
$N$ fractional D2-branes extended along $x^0,x^1,x^2$ and $M$
D6-branes extended along $x^0,x^1,x^2,x^6,\ldots,x^9\,$, as shown
schematically in the following table, where the symbols -- and $\cdot$
denote respectively coordinates which are longitudinal and transverse
to the branes:
\begin{center}
\begin{tabular}{|c|c|c|c|c|c|c|c|c|c|c|}
\multicolumn{7}{c}{ }&
\multicolumn{4}{c}{$\overbrace{\phantom{\qquad\qquad\quad}}^{\mathbb{R}^4/\mathbb{Z}_2}$}\\
\hline
&0&1&2&3&4&5&6&7&8&9\\
\hline
D2 &--&--&--&$\cdot$&$\cdot$&$\cdot$&$\cdot$&$\cdot$&$\cdot$&$\cdot$\\
\hline
D6 &--&--&--&$\cdot$&$\cdot$&$\cdot$&--&--&--&--\\
\hline
\end{tabular}
\end{center}
A peculiar feature of the fractional branes transverse to
the orbifold space (as the D2-branes that we are considering) is
that they are stuck at the orbifold fixed point $x^6=x^7=x^8=x^9=0\,$.

The orbifold projection breaks half of the supersymmetry of type IIA
theory, and so does the considered D2/D6-brane system. We are then left with
8 supercharges. Thus, at low energy the theory living on $N$
fractional D2-branes is $\mathcal{N}=4\,$, $D=2+1$ SYM theory with
gauge group $SU(N)$. Moreover, from the point of view of this gauge
theory the strings stretching from the D2 to the D6-branes and
vice-versa make up $M$ hypermultiplets which transform in the
fundamental representation of the gauge group.

In order to describe the above system by means of a supergravity solution,
we have to study how the low-energy fields which appear in the effective
action behave in the background \eqref{orb}.
Our background is characterized by the presence of a 2-form
$\omega_2$, Poincar\'e dual to the exceptional 2-cycle
$\Sigma_2$ of the ALE space which is obtained by the resolution of the
orbifold. In the orbifold limit, the volume of $\Sigma_2$ vanishes,
but the background value of the integral of $B_2$ on it has to remain finite
in order to define a sensible CFT
\cite{Aspinwall:1995zi,Douglas:1996xg}:
\begin{equation}\label{asp}
   \int_{\Sigma_2}B_2=\frac{\left(2\pi\sqrt{\alpha'}\right)^2}{2}\,.
\end{equation}
The 2-form $\omega_2$ satisfies the following properties:
\begin{equation}\label{omegaprop}
   \omega_2=-\hd{4} \omega_2\,, \qquad \int_{\Sigma_2}\omega_2=1\,,
   \qquad \int_{\text{ALE}}\omega_2\wedge\omega_2=-\frac{1}{2}\,.
\end{equation}
The supergravity fields can have components along the vanishing cycle,
so the following decomposition holds for the NS-NS two-form
and the R-R three-form:
\begin{equation}\label{decomp}
   B_2=\bar{B}_2+b\,\omega_2\,,\qquad
   C_3=\bar{C}_3+A_1\wedge\omega_2\,.
\end{equation}
Since we will be looking for supergravity solutions which represent branes
without a $B_2$ field in their world-volume, 
in the following we will put $\bar{B}_2=0$, so we simply have:
\begin{equation}\label{fieldform}
   B_2=b\,\omega_2\,,\qquad
   C_3=\bar{C}_3+A_1\wedge\omega_2\,,
\end{equation}
where, because of eq.~\eqref{asp}:
\begin{equation}\label{asp2}
   b=\frac{\left(2\pi\sqrt{\alpha'}\right)^2}{2}+\tilde{b}\,,
\end{equation}
and $\tilde{b}$ represents the fluctuation around the background value
of $b$. 
We will sometimes refer to the fields $b$ and $A_1$ in 
eq.~\eqref{fieldform} as ``twisted'' fields because they correspond to the
massless states of the twisted sector of
type IIA string theory on the orbifold.

Having given the main features of the system that we are going to study,
we now turn to the supergravity solution.

\subsection{The supergravity solution}

In this subsection we will discuss the supergravity solution
describing the fractional D2/D6 system that we have introduced in
the previous subsection. The solution is derived in detail in appendix
\ref{details}. Here we will only summarize the procedure followed to
find it.

The first step is to substitute the decompositions for
$B_2$ and $C_3$ given in eq.~\eqref{fieldform}
into the type IIA supergravity action, and to derive
the equations of motion for the ``untwisted'' fields
$G_{\mu\nu}\,$, $\phi\,$, $\bar{C}_3$ and $C_1$ and for the
``twisted'' ones $b\,$ and $A_1\,$.

Then, we impose the standard Ansatz for the ``untwisted'' fields
corresponding to a D2/D6 system\footnote{The
coordinates are labeled as:
$\alpha,\beta=\left\{0,1,2\right\}\,$, 
$i,j=\left\{3,4,5\right\}$ and
$p,q=\left\{6,7,8,9\right\}\,$.}:
\begin{subequations}
\begin{align}
   ds^2&=H_2^{-5/8}H_6^{-1/8}\eta_{\alpha\beta}dx^\alpha dx^\beta
       +H_2^{3/8}H_6^{7/8}\delta_{ij}dx^idx^j
       +H_2^{3/8}H_6^{-1/8}\delta_{pq}dx^pdx^q\,,\\
   e^\phi&=H_2^{1/4}H_6^{-3/4}\,,\\
   \bar{C}_3&=\left(H_2^{-1}-1\right)dx^0\wedge dx^1\wedge dx^2\,,
\end{align}
\end{subequations}
where the function $H_2$ depends on the radial coordinate $\rho=\sqrt{(x^3)^2+\ldots+(x^9)^2}$
of the space transverse to the D2-brane, while the function
$H_6$ depends only on the radial coordinate 
of the common transverse space $r=\sqrt{\delta_{ij}x^ix^j}\,$.

In order to write down a sensible Ansatz for the fields
$A_1$ and $C_1\,$, we need to take into account
the contribution coming from the boundary action
describing the world-volume theory of the branes.
After some calculation it is easy to get  convinced that the 
following\footnote{When
$M=0\,$, the Ans\"atze for $A_1$ and $C_1$ coincide
with the ones given in Ref.~\cite{Cvetic:2000mh}.} is
a sensible Ansatz for the fields $A_1$ and $C_1\,$:
\begin{subequations}\label{duansC1t}
\begin{align}
  dA_1&=C_1\wedge db+\frac{1}{2}\varepsilon_{ijk}H_6\partial_ib dx^j\wedge dx^k\,,\\
  dC_1&=\frac{1}{2}\varepsilon_{ijk}\partial_iH_6 dx^j\wedge dx^k\,,
 \end{align}
\end{subequations}
where $\varepsilon_{345}=\varepsilon^{345}=+1\,$. 

Substituting our Ans\"atze in the equations of motion and
computing all the relevant contributions coming
from the boundary action $S_b\,$, the final solution for the fractional D2/D6 system
can be expressed in the following form\footnote{Notice
that, in order to easily express the fields $A_1$ and $C_1$ in
eq.s~\eqref{solution}, we have changed coordinates in the common transverse
space into polar coordinates: $(x^3,x^4,x^5)\longrightarrow (r,\theta,\varphi)\,$.}:
\begin{subequations}\label{solution}
\begin{align}
   ds^2&=H_2^{-5/8}H_6^{-1/8}\eta_{\alpha\beta}dx^\alpha dx^\beta
       +H_2^{3/8}H_6^{7/8}\delta_{ij}dx^idx^j
       +H_2^{3/8}H_6^{-1/8}\delta_{pq}dx^pdx^q\,,\\
   e^\phi&=H_2^{1/4}H_6^{-3/4}\,,\\
   \bar{C}_3&=\left(H_2^{-1}-1\right)dx^0\wedge dx^1\wedge dx^2\,,\\
   C_1&=\frac{g_s\sqrt{\alpha'}M}{2}\cos\theta d\varphi\,,\\
   A_1&=-\pi^2\alpha'\frac{g_s\sqrt{\alpha'}(4N-M)}
      {H_6} \cos\theta d\varphi\,,\\
   b&=\frac{Z}{H_6}\,,
 \end{align}
\end{subequations} 
where:
\begin{equation}
   H_6(r)=1+\frac{g_s\sqrt{\alpha'}M}{2r}\,,\qquad
   Z(r)=\frac{\left(2\pi\sqrt{\alpha'}\right)^2}{2}
      \left(1-\frac{g_s\sqrt{\alpha'}(2N-M)}{r}\right)\,,
\end{equation}
and where $H_2$ is the solution of the following equation
(see eq.~\eqref{H2} in appendix \ref{details}):
\begin{equation}
   \left(\delta^{ij}\partial_i\partial_j 
      +H_6\delta^{pq}\partial_p\partial_q\right) H_2
      +\frac{1}{2}H_6\delta^{ij}\partial_i b\partial_j b\delta(x^6)\cdots\delta(x^9)
      +\kappa T_2N\delta(x^3)\cdots\delta(x^9)=0\,.
\end{equation}

{From} the solution given in eq.~\eqref{solution} we can also compute
the expressions for the fields $C_7$ and $A_3$ which appear naturally 
in the string theory.
The duality relations are\footnote{The duality relations can be derived
from the equations of motion \eqref{orbeqC1} and \eqref{orbeqA1}, for which the terms
coming from the boundary action vanish (see appendix \ref{details}).}:
\begin{subequations}
\begin{align}
  dC_7 &= -e^{3\phi/2} \hd{} dC_1\,,\\
  dA_3 &= e^{\phi/2}\ \hd{6} G_2 - db \wedge \bar{C}_3\,,
\end{align}
\end{subequations}
and the explicit computation gives:
\begin{subequations}
\begin{align}
  C_7 &= (H_6^{-1}-1) \ dx^0\wedge\cdots\wedge dx^6\,,\\
  A_3 &= \tilde{b} \  dx^0\wedge dx^1\wedge dx^2\,.
\end{align}
\end{subequations}
Notice that the field $C_7$ has a quite standard expression, due to the
specific form of the Ansatz in eq.~\eqref{duansC1t}.

\subsection{Probing the fractional brane solution}\label{gaugetheory}

In this section we will study the world-volume theory of a probe
fractional D2-brane, which is placed in the background given in 
eq.s~\eqref{solution} 
at some finite distance $r$ in the transverse space 
$\left\{x^3,x^4,x^5\right\}$. This will give us 
information about the Coulomb branch 
of $\mathcal{N}=4\,$ three-dimensional super Yang--Mills theory with
gauge group $SU(N+1)$ broken into $ SU(N)\times U(1)$, coupled to $M$ 
hypermultiplets in the fundamental representation of the 
gauge group. 

Let us start from the world-volume action for a single fractional
D2-brane, which is given by eq.~\eqref{FracDp} 
in the case of $p=2\,$: 
\begin{multline}\label{probact}
   S_{\text{probe}}=-\frac{T_{2}}{2\kappa}
         \int{d^{3}\xi e^{-\phi/4}
         \sqrt{-\det \big[G_{\alpha\beta}+e^{-\phi/2}
         2\pi\alpha'F_{\alpha\beta}\big]} 
         \left(1+\frac{\tilde{b}}{2\pi^2\alpha'}\right)}  \\
      +\frac{T_{2}}{2\kappa}\int_{{\cal M}_3}{\big(
         {\cal C}_3+2\pi\alpha'{\cal C}_1\wedge F\big)}\,, 
\end{multline}
where all bulk fields are understood to be
pullbacks onto the brane world-volume and the
fields $\mathcal{C}_3$ and $\mathcal{C}_1$ are given by:
\begin{subequations}\label{curlyc}
\begin{align}
    \mathcal{C}_3&=\bar{C}_3\left(1+\frac{\tilde{b}}{2\pi^2\alpha'}\right)+
         \frac{A_3}{2\pi^2\alpha'}=\frac{1}{2\pi^2\alpha'}\frac{Z}{H_2H_6}-1\,,\\
    \mathcal{C}_1&={C}_1\left(1+\frac{\tilde{b}}{2\pi^2\alpha'}\right)+
         \frac{A_1}{2\pi^2\alpha'}=-g_s\sqrt{\alpha'}(2N-M)\cos\theta d\varphi\,.
\end{align}
\end{subequations}

The computation is analogous to the one performed in section
\ref{wprobesect}. We fix the static gauge, and regard the coordinates
$\{x^3,x^4,x^5\}$ transverse to the probe brane as Higgs fields of the
dual gauge theory: $x^i=2\pi\alpha'\Phi^i$. We also define polar
coordinates $(\mu,\theta,\varphi)$ in the moduli space of the
$\Phi^i$, so that $r=2\pi\alpha'\mu\,$.

Expanding the DBI action for slowly varying world-volume fields
and keeping only up to quadratic terms in their derivatives we get: 
\begin{multline}\label{probDBI}
    S_{\text{DBI}}\simeq -\frac{\sqrt{\alpha'}}{2g_s}
     \int{d^3 x \frac{Z}{2\pi^2\alpha'}
     \left(\frac{1}{2}
     \eta^{\alpha\beta}\delta_{ij}\partial_\alpha\Phi^i\partial_\beta\Phi^j+
     \frac{1}{4}\eta^{\alpha\beta}\eta^{\gamma\delta}
     F_{\alpha\gamma}F_{\beta\delta}\right)} \\ 
     -\frac{T_2}{2\kappa}
     \int{d^3x \frac{1}{2\pi^2\alpha'}\frac{Z}{H_2H_6}}\,.
\end{multline}
Turning to the WZ part and substituting the expressions \eqref{curlyc}
into eq.~\eqref{probact} we get:
\begin{equation}\label{probWZ}
   S_{\text{WZ}}=\frac{T_2}{2\kappa}\left[
     \int{d^3x \left(\frac{1}{2\pi^2\alpha'}\frac{Z}{H_2H_6}-1\right)}
     +\int_{\mathcal{M}_3}2\pi\alpha'\mathcal{C}_\varphi d\varphi\wedge F\right]\,.
\end{equation}
We easily see that the position-dependent terms cancel as expected
because fractional D2-branes are BPS states and do not interact with
the D2/D6 system. Ignoring the constant
potential, the final result is:
\begin{multline}\label{notsofinalprobe}
    S_{\text{probe}}= -\frac{\sqrt{\alpha'}}{4g_s}
     \int d^3 x \frac{Z}{2\pi^2\alpha'}
     \left\{\frac{1}{2}
     \big[(\partial\mu)^2+\mu^2\left((\partial\theta)^2+\sin^2\theta(\partial\varphi)^2\right)\big]
     +\frac{1}{4}F^2\right\} \\
     -\frac{1}{16\pi}\int d^3 x (2N-M)\cos\theta 
     \varepsilon^{\alpha\beta\gamma}\partial_\alpha\varphi F_{\beta\gamma}\,.
\end{multline} 
When $Z=0\,$ the effective tension of the probe vanishes and this
means that also in this case, as expected, an enhan\c con mechanism is
taking place at the radius:
\begin{equation}
   r_{\text{e}}=\sqrt{\alpha'}g_s(2N-M)\,.
\end{equation}
Substituting in \eqref{notsofinalprobe}
the expression of $Z$ in terms of $\mu\,$,
we obtain:
\begin{multline}\label{finalprobe}
    S_{\text{probe}}= -\frac{\sqrt{\alpha'}}{4g_s}
     \int d^3 x \left[1-\frac{g_s(2N-M)}{2\pi\sqrt{\alpha'}\mu}\right]
     \left\{\frac{1}{2}
     \big[(\partial\mu)^2+\mu^2\left((\partial\theta)^2+\sin^2\theta(\partial\varphi)^2\right)\big]
     +\frac{1}{4}F^2\right\} \\
     -\frac{1}{16\pi}\int d^3 x (2N-M)\cos\theta 
     \varepsilon^{\alpha\beta\gamma}\partial_\alpha\varphi F_{\beta\gamma}\,.
\end{multline} 
{From} the coefficient
of the gauge field kinetic term in the previous action we can read the
running coupling constant:
\begin{equation}\label{running}
   \frac{1}{g^2_{\text{YM}}(\mu)}=
   \frac{1}{g^2_{\text{YM}}}
   \left(1-g^2_{\text{YM}}\frac{2N-M}{8\pi\mu}\right)\, ,
\end{equation}
where we have defined the bare coupling as:
\begin{equation}\label{bare}
   g^2_{\text{YM}}=\frac{4g_s}{\sqrt{\alpha'}}\,.
\end{equation}
Eq.~\eqref{running} is exactly what expected for the gauge theory
under consideration, as shown in appendix \ref{gaugerun}.

Exactly as in the case of the wrapped branes described in section
\ref{wrappedD4}, eq.~\eqref{finalprobe} does not give explicitly the 
full hyperK\"ahler metric
on the moduli space of the gauge theory.
We can obtain the needed extra modulus by
dualising the vector field into a scalar,
using exactly the same procedure which brought us from eq.~\eqref{wfinprobe}
to eq.~\eqref{wMetricMS} in section \ref{wrappedD4}.
The final result is\footnote{In this case, the
dualisation procedure can also be done
directly in the original three-dimensional world-volume 
action, as in Ref.s~\cite{Schmidhuber:1996,Johnson:2000qt}. The
result that one obtains coincides with that in eq.~\eqref{MetricMS}.}:
\begin{equation}\label{MetricMS}
   ds^2_{\mathcal{M}}=
     \frac{1}{g^2_{\text{YM}}(\mu)}
     \left(d\mu^2+\mu^2d\Omega^2\right)
     +g^2_{\text{YM}}(\mu)
     \left(d\Sigma+
     \frac{(2N-M)\cos\theta}{8\pi}d\varphi\right)^2\,,
\end{equation}
where $d\Omega^2=d\theta^2+\sin^2\theta d\varphi^2\,$.
If we put $M=0$ this metric coincides with the one found
in eq.~\eqref{wMetricMS} by probing the geometry of $N$
D4-branes wrapped on $S^2\,$. Again, we have found
the hyperK\"ahler Taub-NUT metric,
but with a ``negative mass'' which makes it singular. Also in this
case we have only recovered  the perturbative behaviour
of the gauge theory.

\subsubsection*{Complete action for a D6-brane extended along the orbifold}

As in the case of the D7-brane analyzed in
Ref.~\cite{Bertolini:2001qa}, the supergravity 
solution corresponding to our D2/D6 system can also be used to get the 
form of 
the complete world-volume action for a D6-brane
extended along the whole orbifold space. In fact, in deriving the
classical solution corresponding to the D2/D6 system it was enough to
consider only the linear terms in the bulk fields. However, since the
D2/D6 system is BPS, we expect that when we insert the corresponding
classical solution into the
world-volume action of either the D2-brane or the D6-brane, we
obtain a constant result. In the previous subsection we have seen that
this is the case for a fractional D2-brane. If we instead insert the
classical solution into the action of a D6-brane given by
(as it follows from eq.~\eqref{D6act} for $p=2$):
\begin{multline}
   S_{6}=\frac{T_{6}}{\kappa}\left\{
           -\int d^{7}\xi e^{\frac{3}{4}\phi}\sqrt{-\det G_{\rho\sigma}}
           +\int_{\mathcal{M}_{7}}C_{7}\right\}\\
           +\frac{T_2}{2\kappa}\frac{1}{2(2\pi\sqrt{\alpha'})^2}\left\{
           \int d^{3}\xi \sqrt{-\det G_{\alpha\beta}}\ \tilde{b}
           -\int_{\mathcal{M}_{3}}A_{3}\right\}+\ldots\,,
\end{multline} 
we get terms that are dependent on the distance $r$ between the
D6-brane and the system D2/D6 described by the classical solution.
The situation here is exactly the
same as the one found in Ref.~\cite{Bertolini:2001qa}, and as in that case
we must add to the previous action higher order terms that restore
the no-force condition. Including them we arrive at the following
boundary action for a D6-brane extended along the whole orbifold space:
\begin{equation}
\begin{split}
   S_6&=\frac{T_6}{\kappa}\left\{
           -\int d^{7}x e^{\frac{3}{4}\phi}\sqrt{-\det G_{\rho\sigma}}
           +\int_{\mathcal{M}_{7}}C_{7}\right\}\\
           &+\frac{T_2}{2\kappa}\frac{1}{2(2\pi\sqrt{\alpha'})^2}\Bigg\{
           \int d^{3}\xi e^{-\phi/4}\sqrt{-\det G_{\alpha\beta}}\ 
           \tilde{b}\left(1+\frac{\tilde{b}}{4\pi^2\alpha'}\right)\\
           &-\int_{\mathcal{M}_{3}}A_{3}\left(1+\frac{\tilde{b}}{4\pi^2\alpha'}\right)
           -\int_{\mathcal{M}_{3}}\bar{C}_{3} \tilde{b}
           \left(1+\frac{\tilde{b}}{4\pi^2\alpha'}\right)\
           \Bigg\}\,.
\end{split}
\end{equation}

\section{Discussion and conclusions}\label{discussion}

In this final section we want to compare the two approaches that we
have described in the previous section. Let us compare the two
cases dual to the pure gauge theory, that is, in absence of
D6-branes ($M=0$) in the fractional brane case.

Let us start by summarizing some of the properties and differences of 
the two systems:

\begin{itemize}
\item Both systems are able to capture only the perturbative
dynamics of $\mathcal{N}=4\,$, $D=2+1$ SYM theory.
\item The role of the running coupling constant is played in the
two supergravity solutions by two parameters: the ``running
volume'' $\mathcal{Z}$ of the 2-cycle for the wrapped D4-branes and
the ``twisted $B$-field'' $b$ for the fractional D2-branes.
\item The enhan\c con, where the gauge coupling $g_{\text{YM}}$ 
diverges, is located at the locus where respectively $\mathcal{Z}=0$ 
and $b=0\,$.
\end{itemize}

Does it exist a closer relationship between the two setups? Both
systems consist of wrapped branes. In fact, on the one hand, as we
have seen in section \ref{fractionalsystem}, a fractional D$p$-brane
can be seen as a D$(p+2)$-brane wrapped on the vanishing two-cycle of
the ALE space which corresponds to the blow-up of the orbifold.  On
the other hand we have also seen that the D4-branes considered in
section \ref{wrappedD4} are wrapped on a two-sphere inside the same
ALE space, as explicitly shown in eq.~\eqref{ALE}. In fact, the two
systems provide exactly the same information about the gauge theory
living on their world-volume. In order to see the connection between
the two systems it is useful to write down a general formula that
provides the perturbative running coupling constant of a general
$(p+1)$-dimensional gauge theory living on the flat part of the
world-volume of a D$(p+2)$-brane wrapped on a (vanishing or not)
two-cycle $\Sigma_2$.  It is given by the following expression:
\begin{equation}\label{masterformula}
   \frac{1}{g^2_{\text{YM}}(\mu)}=\frac{V_{\text{ST}}\left(\Sigma_2\right)}{g^2_{\text{D}p}}\,,
\end{equation}
where $g^2_{\text{D}p}=2(2\pi)^{p-2}g_s{\alpha'}^{(p-3)/2}$ is
the usual (bare) coupling constant of the gauge theory living
on a D$p$-brane in flat space and the dimensionless
``stringy volume'' $V_{\text{ST}}$ is given
by:
\begin{equation}
V_{\text{ST}}\left(\Sigma_2\right)=
\frac{1}{\left(2\pi\sqrt{\alpha'}\right)^2}
\int d^2\zeta \sqrt{-\det\left(\mathcal{G}_{AB}+
B_{AB}\right)}\,,   
\label{volume45}
\end{equation}
where $\zeta^{1,2}$ parameterizes the cycle $\Sigma_2\,$, while
$\mathcal{G}_{AB}$ and $B_{AB}$ ($A,B=1,2\,$) are the bulk metric
\emph{without any warp factors} and the $B$-field on the cycle.

We can easily see that the formula in eq.~\eqref{masterformula} holds
for the systems considered in sections \ref{wrappedD4} and \ref{fractionalsystem}. For the
three-dimensional gauge theory at hand, we have
$g^2_{\text{D}2}=\frac{2g_s}{\sqrt{\alpha'}}\,$. 

For the case of
the D4-branes wrapped on $S^2\,$, we have:
\begin{equation}
   \mathcal{G}^{\text{w}}=\mathcal{Z}R_0^2\begin{pmatrix}1&0\\0&\sin^2\ttheta\end{pmatrix}\,,\qquad
   B^{\text{w}}=0\,,
\end{equation}
so that the ``stringy volume'' is given by:
\begin{equation}\label{VSTwrap}
   V_{\text{ST}}^{\text{w}}\left(\Sigma_2\right)=\frac{\mathcal{Z}R_0^2}{\pi^2\alpha'}\,.
\end{equation}
Substituting eq.~\eqref{VSTwrap} in eq.~\eqref{masterformula}
we get the correct running for $\mathcal{N}=4\,$, $D=2+1$ SYM theory
with $SU(N)$ gauge group:
\begin{equation}
   \frac{1}{(g_{\text{YM}}^{\text{w}})^2(\mu)}=
   \frac{1}{(g_{\text{YM}}^{\text{w}})^2}
   \left(1-\frac{(g_{\text{YM}}^{\text{w}})^2 N}{4\pi\mu}\right)\,,
\end{equation}
where the bare coupling is defined as in eq.~\eqref{wbare} as follows:
$(g_{\text{YM}}^{\text{w}})^2=\frac{2\pi g_s \sqrt{\alpha'}}{R_0^2}\,$.

If instead we consider the fractional D2-branes as D4-branes wrapped on
the vanishing cycle $\Sigma_2\,$, we find:
\begin{equation}
   \mathcal{G}^{\text{f}}=0\,,\qquad
   B^{\text{f}}=Z\omega_2\,.
\end{equation}
Now the ``stringy volume'' is:
\begin{equation}\label{VSTfrac}
   V_{\text{ST}}^{\text{f}}\left(\Sigma_2\right)=\frac{Z}{4\pi^2\alpha'}\,.
\end{equation}
and substituting eq.~\eqref{VSTfrac} in eq.~\eqref{masterformula}
we get again the correct running coupling constant:
\begin{equation}
   \frac{1}{(g_{\text{YM}}^{\text{f}})^2(\mu)}=
   \frac{1}{(g_{\text{YM}}^{\text{f}})^2}
   \left(1-\frac{(g_{\text{YM}}^{\text{f}})^2 N}{4\pi\mu}\right)\,.
\end{equation}
where now the bare coupling constant is defined as
$(g_{\text{YM}}^{\text{f}})^2=\frac{4g_s}{\sqrt{\alpha'}}\,$
as in eq.~\eqref{bare}.

Notice also that if we choose the ``background value'' $V_0$ of the
geometrical volume on which the D4-branes are wrapped (that is, the
volume of the two-sphere inside the Eguchi--Hanson space in
eq.~\eqref{ALE}, once we remove the branes setting
$H=\widetilde{\mathcal{Z}}=1$) in such a way that it coincides with
the background value of the $B$-field of the fractional brane case,
$V_0=4\pi R_0^2=\tfrac{(2\pi\sqrt{\alpha'})^2}{2}\,$, the bare
coupling constants (and enhan\c con radii) computed in the two cases
become exactly the same in terms of string parameters.

One can see that the formula \eqref{masterformula} works
perfectly also, for instance, for the case of $\mathcal{N}=2$ SYM
in four dimensions, applying it to the fractional D3-brane solution
of Ref.s~\cite{Bertolini:2001dk,Polchinski:2001mx}
and to the wrapped D5-brane solution of Ref.~\cite{Gauntlett:2001ps}.

Although the two systems give the same perturbative gauge coupling
constant, there seems not to be a ``physical'' limit in which one can
obtain the fractional brane solution from the wrapped one or
vice-versa, playing with the volume of the cycle. 
This is due to the fact that the two Ans\"atze are
radically different in the warp factors, which are respectively
the ones of a D4 and of a D2-brane, and is also due to the absence of a $B$-field
on the world-volume of the D4-branes wrapped on $S^2$. On the other hand,
if we look at the whole moduli space of the four-dimensional ALE
space, we see that it is characterized by the volume
of the exceptional cycle and by the flux of the $B$-field on it. These
are the two moduli that are combined into the ``stringy volume'' in
eq.~(\ref{volume45}) which, as we have seen, provides the running
coupling constant. The situation in the two cases
can then be summarized by the following diagram:
{\small
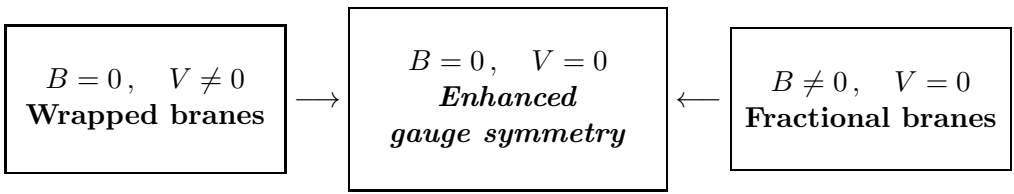
\begin{displaymath}
\boxed{\parbox{3.5cm}{\begin{center}$B=0\,,\quad V\neq0$\\ {\bf Wrapped branes}\end{center}}} 
\longrightarrow
\boxed{\parbox{4cm}{\begin{center}$B=0\,,\quad V=0$\\ {\bf \emph{Enhanced}}\\{\bf \emph{gauge symmetry}}\end{center}}} 
\longleftarrow
\boxed{\parbox{3.5cm}{\begin{center}$B\neq0\,,\quad V=0$\\ {\bf Fractional branes}\end{center}}} 
\end{displaymath}}
where, in the case of the wrapped D4-branes,
we are keeping the size of the cycle finite and the $B$-flux
vanishing, while in the case of the fractional D2-branes 
the geometrical volume shrinks to zero size and in order to have a 
\emph{conformal} orbifold we must give a definite fixed background 
value to the $B$-flux, which in the case of
the $\mathbb{Z}_2$ orbifold is $\tfrac{(2\pi\sqrt{\alpha'})^2}{2}\,$. 

The limiting case in which both the geometrical volume of the cycle
and the $B$-flux are taken to vanish is the point where the theory on
the Calabi--Yau space manifests an enhanced gauge symmetry, which is
at the origin of the enchan\c con mechanism. This is the point where
the ``stringy volume'' vanishes and the supergravity solutions break
down.
 
\subsection*{Acknowledgements}

We would like to thank M.~Bertolini, M.~Frau, A.~Lerda, P.~Merlatti
and T.~Ort\'{\i}n for enlightening discussions. H.E. wishes to thank
NORDITA for kind hospitality. The work of E.I. and E.L.--T. is
supported in part by a European Commission Marie Curie Training Site
Fellowship under the Contract No.~HPMT-CT-2000-00010.

\appendix

\section{Finding the supergravity solutions}\label{appsugrasol}

In the two sections of this appendix we will describe the way
in which we have obtained the supergravity solutions describing,
respectively, D4-branes wrapped on $S^2$ and fractional D2/D6-branes on the
orbifold $\mathbb{R}^4/\mathbb{Z}_2\,$.

We will be using the following conventions:
\begin{itemize}
\item A metric in $D$ dimensions has signature $(-,+^{D-1})\,$.
\item $\varepsilon$-symbols in $D$ dimensions are defined in such a way that\\
$\varepsilon^{012\dotsm (D-1)}=-\varepsilon_{012\dotsm (D-1)}=+1\,$.
\item A $p$ form is defined as
$\omega_p=\frac{1}{p!}\omega_{\mu_1 \dotsm\mu_p}dx^{\mu_1}\wedge\dotsm\wedge dx^{\mu_p}\,$.
\item The Hodge dual $\hd{D}$ in $D$ dimensions is defined as\\
$\hd{D}\omega_p=\frac{\sqrt{-\det G_D}}{p!(D-p)!}
\varepsilon_{\nu_1\dotsm \nu_{D-p}\mu_1 \dotsm\mu_p}
\omega^{\mu_1\dotsm\mu_p}
dx^{\nu_1}\wedge\dotsm\wedge dx^{\nu_{D-p}}\,$.\\
Moreover, $\hd{}$ denotes $\hd{10}$ and $\hde$ denotes $\hd{11}\,$.
\end{itemize}

\subsection{D4-branes wrapped on $S^2$}\label{D4onS2SugraSolution}

In this appendix we explain how we have obtained the type IIA
supergravity solution describing $N$ D4-branes wrapped on $S^2\,$,
using the techniques and the solutions given in Ref.~\cite{MN}.

Our procedure will be the following. We want to obtain the solution
for the D4-branes by compactifying the solution for the M5-branes
wrapped on $S^2\,$. The latter is found by uplifting to eleven
dimensions a solution of 7-dimensional gauged supergravity with the
correct identification between spin connection and gauge connection,
by means of the formulas given in Ref.~\cite{Cv1}.

\subsubsection*{The seven dimensional solution}

The starting point is the seven dimensional gauged supergravity 
considered in Ref.~\cite{MN}. Following that paper, we consider
a $U(1)\times U(1)$ consistent truncation of the $SO(5)$ gauged 
supergravity arising when one compactifies eleven dimensional 
supergravity on $S^4$. The bosonic field content
of the truncated theory consists of two $U(1)$ gauge fields ($A^{(1,2)}$), 
two scalar fields ($\lambda_{1,2}$) and a metric. 

The full solution of seven dimensional 
gauged supergravity is:
\begin{subequations}
\label{7dimsolution}
\begin{align}
  ds_{(7)}^2&=\left(\frac{R_A}{R_0}\right)^2e^{2\rho}e^{\lambda}\, 
            \eta_{ij}d\xi^id\xi^j 
          +R_A^2\left(e^\lambda(e^{2\rho}
          -{\textstyle\frac{1}{4}}) (d\ttheta^2+\sin^2\ttheta d\tphi^2)+
          e^{-4\lambda}d\rho^2\right)\, , \\
  A^{(1)}&=\frac{R_A}{4}\cos\ttheta d\tphi\, , \ \ A^{(2)}=0\, , \\
  \lambda&\equiv\lambda_2\, , \ \ 2\lambda_1+3\lambda_2=0\, , \\ 
  e^{5\lambda}&=\frac{e^{2\rho}+ke^{-2\rho}-\frac{1}{2}}
                {e^{2\rho}-\frac{1}{4}}\, , 
\end{align}
\end{subequations}
This solution is exactly the one given in Ref.~\cite{MN}, although we
have kept track of units and we have used standard spherical
coordinates $\ttheta$ and $\tphi$ for the two-sphere on which the
branes are wrapped.  $R_A=2(\pi N)^{1/3}l_p$ is the radius of the
$AdS_7$ space appearing in the near horizon limit of the usual flat
M5-brane solution, and $R_0$ is an arbitrary integration constant with
dimension of a length (which is $(C_2)^{-1/2}$ of eq.~(24) in
Ref.~\cite{MN}). Finally, $k$ is a (dimensionless) integration
constant, which was called $C_1$ in Ref.~\cite{MN}. All the
coordinates entering in the above solution are dimensionless, except
those spanning the unwrapped part of the world-volume of the brane,
$\xi^i\, , i=0,\ldots,3$, which have dimensions of a length.

\subsubsection*{Uplift formulas and eleven dimensional solution}

The seven-dimensional solution can be lifted to eleven dimensions 
with the help of eq.s~(4.1) and (4.2) of Ref.~\cite{Cv1}, which we 
rewrite here:
\begin{subequations}
\label{uplift}
\begin{align}
  d\hat{s}^2&=\tilde{\Delta}^{1/3}ds_{(7)}^2+g^{-2}\tilde{\Delta}^{-2/3}
   \left(X_0^{-1}d\mu_0^2+
   \sum_{i=1}^{2}
   X_i^{-1}\left(d\mu_i^2+\mu_i^2\left(d\phi_i+g{\cal A}^{(i)}
   \right)^2\right)\right)\, , \\ 
  \hde d\hat{C}_3&=
   2g\sum_{\alpha=0}^{2}\left(X_\alpha^2\mu_\alpha^2-
    \tilde{\Delta}X_\alpha\right)\varepsilon_{(7)}+
   g\tilde{\Delta}X_0\, \varepsilon_{(7)}+
   \frac{1}{2g}\sum_{\alpha=0}^{2}X_\alpha^{-1}\ \hd{7}dX_\alpha
    \wedge d(\mu_\alpha^2) \nonumber \\
   &\ \ +\frac{1}{2g^2}\sum_{i=1}^{2}
   X_i^{-2}d(\mu_i^2)\wedge\left(d\phi_i+g{\cal A}^{(i)}\right)
   \wedge\hd{7}{\cal F}^{(i)}\, .
\end{align}
\end{subequations}
Here and below, hats will always refer to eleven-dimensional quantities.
The above formulas are written in the notation of Ref.~\cite{Cv1}: 
$g$ is the seven 
dimensional gauged supergravity coupling constant, $\varepsilon_{(7)}$ 
is the seven dimensional volume form, ${\cal A}^{(1,2)}$ are the two
$U(1)$ gauge fields, the $X_{\alpha}$ are a suitable parameterization of
the 2 scalars present in the theory and $\tilde{\Delta}$ is given by:
\begin{displaymath}
  \tilde{\Delta}\equiv\sum_{\alpha=0}^{2}X_\alpha\mu_\alpha^2\, , 
\end{displaymath}
where $\mu_\alpha$ parameterize a two-sphere:
$\mu_0^2+\mu_1^2+\mu_2^2=1$. The quantities appearing in the uplift
formulas are given in terms of those appearing in
eq.~(\ref{7dimsolution}) by the following expressions:
\begin{equation}
\label{translation}
\begin{aligned}
  \displaystyle{\frac{1}{g^2}}&=
    \displaystyle{\left(\frac{R_A}{2}\right)^2}\, , \\
  X_0&=X_1=e^{2\lambda}\, , \\
  X_2&=e^{-3\lambda}\, , \\
  {\cal A}^{(1,2)}&=2A^{(2,1)}\, , \\ 
  \Delta&\equiv e^{3\lambda}\tilde{\Delta}=
    e^{5\lambda}\cos^2\chi+\sin^2\chi\, , \\ 
  \varepsilon_{(7)}&=-\sqrt{-\det G_{(7)}}\, 
    d\xi_0\wedge \cdots d\xi_3\wedge d\ttheta\wedge d\tphi\wedge d\rho\, , 
\end{aligned} 
\end{equation}
where we have chosen the following parameterization for $\mu_i\,$: 
\begin{equation}
\begin{aligned}
  \mu_0&=\cos\chi\cos\theta\, , \\ 
  \mu_1&=\cos\chi\sin\theta\, , \\ 
  \mu_2&=\sin\chi\, . 
\end{aligned}
\end{equation}
By using the above expressions we are now ready to write the full 
solution in eleven dimensions:
\begin{subequations}
\label{11dimsolution}
\begin{align}
  d\hat{s}^2&=\Delta^{1/3}\left(\left(\frac{R_A}{R_0}\right)^2
    e^{2\rho}\, \eta_{ij}d\xi^id\xi^j + 
    R_A^2(e^{2\rho}-{\textstyle\frac{1}{4}})
    (d\ttheta^2+\sin^2\ttheta d\tphi^2)\right) \nonumber \\ 
   &+\Delta^{-2/3}\left(\frac{R_A}{2}\right)^2\left(
    \frac{4\Delta}{e^{5\lambda}}d\rho^2+\Delta d\chi^2+
    \cos^2\chi(d\theta^2+\sin^2\theta d\varphi^2) +   
    e^{5\lambda}\sin^2\chi\left(d\psi+\cos\ttheta d\tphi\right)^2 \right)\, , \\ 
  \hde d\hat{C}_3&=\frac{R_A^6}{R_0^4}\, 
    e^{4\rho}(e^{2\rho}-\tfrac{1}{4})\sin\ttheta\Bigg(
    2(\Delta+2)\, d\xi^0\wedge\cdots\wedge d\xi^3
    \wedge d\ttheta\wedge d\tphi\wedge d\rho  \nonumber \\
  &\ \ \ +\frac{1}{4}\partial_\rho(e^{5\lambda})\sin(2\chi)\, 
    d\xi^0\wedge\cdots\wedge d\xi^3
    \wedge d\ttheta\wedge d\tphi\wedge d\chi \nonumber \\ 
  &\ \ \ +\frac{1}{16}\frac{e^{5\lambda}\sin(2\chi)}
    {(e^{2\rho}-\frac{1}{4})^2\sin\ttheta}\, 
    d\xi^0\wedge\cdots\wedge d\xi^3
    \wedge d\rho \wedge d\chi \wedge \left(d\psi+\cos\ttheta d\tphi\right)\Bigg)\,  
  \label{11dimsolutionb} 
\end{align}
\end{subequations}
where we have also relabeled the angles
appearing in~(\ref{uplift}): $\phi_1=\varphi\, , \phi_2=\psi\,$.
This solution describes the near
horizon geometry of an M5-brane wrapped on a two-sphere. The unwrapped
world-volume coordinates are $\xi^0, ..., \xi^3$, the wrapped ones are
$\ttheta$ and $\tphi$, and the remaining coordinates are transverse to the
M5. It can be easily seen that the metric in eq.~(\ref{11dimsolution})
reduces to the one given in eq.~(26) of Ref.~\cite{MN} if we restrict
ourselves to work at the IR fixed point analysed there.

From eq.~(\ref{11dimsolutionb}) we can compute the three-form
potential. It is equal to:
\begin{equation}
\label{C3}
\begin{aligned}
  \hat{C}_3=&\frac{R_A^3}{8}\, \frac{e^{5\lambda}\cos^3\chi\cos\theta\sin\ttheta}
        {\Delta}\, d\ttheta\wedge d\tphi\wedge d\varphi  \\ 
      &+\frac{R_A^3}{8}\, \frac{e^{5\lambda}
        (\Delta+2)\cos^2\chi\sin\chi\cos\theta}{\Delta^2}\, 
        d\chi\wedge d\varphi\wedge\left(d\psi+\cos\ttheta d\tphi\right) \\
      &+\frac{R_A^3}{8}\, \frac{\partial_\rho(e^{5\lambda})
        \cos^3\chi\sin^2\chi\cos\theta}{\Delta^2}\, 
        d\rho\wedge d\varphi\wedge\left(d\psi+\cos\ttheta d\tphi\right)\, . \\ 
\end{aligned}
\end{equation}
The last step consists in compactifying to ten dimensions the M5-brane 
solution just obtained along one of its non-wrapped world-volume 
coordinates (that we choose to be $\xi^3$) to get the solution
describing the geometry 
of $N$ wrapped D4-branes.
The compactification is obtained by means of the standard expressions in 
the ten dimensional string frame:
\begin{subequations}
\label{compactification}
\begin{align}
  (G_{\text{st}})_{\mu\nu}&=(\hat{G}_{33})^{1/2}\hat{G}_{\mu\nu}\, , \\
  e^{2\phi}&=(\hat{G}_{33})^{3/2}\, , \\
  (C_3)_{\mu\nu\rho}&=(\hat{C}_3)_{\mu\nu\rho}\,,  
\end{align}
\end{subequations}
where we have split eleven dimensional indices in 
$\hat{\mu}=\left\{\mu, \xi^3\right\}\,$. This is all we need to get 
the final expression for the wrapped D4-brane solution presented in 
section~\ref{SectTSG-D4onS2}.


\subsection{Fractional D2/D6-brane system}\label{details}

In this section we will describe in some detail how to find the
supergravity solution describing a fractional D2/D6-brane
system. We will always work in the
Einstein frame. The Type IIA effective action in the orbifold
background \eqref{orb} is given, in our conventions, by:
\begin{multline}\label{IIAAction}      
  S_{\text{IIA}}=
            \frac{1}{2\kappa^2}\Bigg\{
            \int d^{10}x\sqrt{-G}\, R-
            \frac{1}{2}
            \int\Big( d\phi\wedge\hd{} d\phi
                 + e^{-\phi}H_3\wedge\hd{} H_3 \\
                 - e^{3\phi/2}F_2\wedge\hd{} F_2   
                 - e^{\phi/2}\tilde{F}_4\wedge\hd{}\tilde{F}_4
                 + B_2 \wedge F_4\wedge F_4     
            \Big)\Bigg\}  \, ,
\end{multline}
where the field strengths are given by:
\begin{equation}
  H_3=dB_2\, , \qquad F_2=dC_1\, , \qquad F_4=dC_3\, , \qquad
  \tilde{F}_4=F_4-C_1\wedge H_3\, , 
\end{equation}
and $\kappa=8\pi^{7/2}g_s{\alpha'}^2$. In order to find a D-brane 
solution we must add to the previous bulk action a boundary action
whose corresponding Lagrangian we call $\mathcal{L}_b\,$. The equations of
motion are then obtained by
varying the total action $S_{\text{IIA}}+S_b\,$.

The first step in order to find a supergravity solution in the
orbifold background (see for instance Ref.~\cite{Bertolini:2001dk}) is
substituting in the action \eqref{IIAAction} the form
\eqref{fieldform} of the fields:
\begin{equation}
   B_2=b\,\omega_2\,,\qquad C_3=\bar{C}_3+A_1\wedge\omega_2\,.
\end{equation}
Recalling that the 2-form $\omega_2$ is normalized as in eq.~\eqref{omegaprop},
one obtains:
\begin{multline}\label{6dimaction}
  S^\prime_{\text{IIA}}=
      \frac{1}{2\kappa^2}\Bigg\{
      \int d^{10}x\sqrt{-G}\, R
      -\frac{1}{2}\int\Big(d\phi\wedge\hd{}d\phi
      - e^{3\phi/2}dC_1\wedge\hd{} dC_1
      - e^{\phi/2}d\bar{C}_3\wedge\hd{}d\bar{C}_3
      \Big)\\
      -\frac{1}{4}\int_{\mathbb{R}^{1,5}}\Big(e^{-\phi}db\wedge\hd{6}db
      -e^{\phi/2}G_2\wedge\hd{6}G_2
      -2b\wedge d\bar{C}_3\wedge dA_1
      \Big)\Bigg\}\,,   
\end{multline}
where we have introduced the quantity: 
\begin{equation}
  G_2\equiv dA_1-C_1\wedge db\, .
\end{equation}
By varying the previous action one finds the equations of motion for
the fields $C_1\,$, $\bar{C}_3\,$, $A_1\,$, $b$ and $\phi$ respectively:
\begin{subequations}\label{orbeqs}
\begin{align} 
   d\left(e^{3\phi/2}\ \hd{} dC_1\right)
      -\frac{1}{2}e^{\phi/2}db\wedge\hd{6}G_2\wedge\Omega_4
      +2\kappa^2\frac{\delta \mathcal{L}_b}{\delta C_1}&=0\,,\label{orbeqC1}\\
   d\left(e^{\phi/2}\ \hd{} d\bar{C}_3\right)
      +\frac{1}{2}db\wedge dA_1\wedge\Omega_4
      +2\kappa^2\frac{\delta \mathcal{L}_b}{\delta \bar{C}_3}&=0\,,\label{orbeqC3}\displaybreak[1]\\
   d\left(e^{\phi/2}\ \hd{6} G_2\right)
      +db\wedge d\bar{C}_3+4\kappa^2\frac{\delta \mathcal{L}_b}{\delta A_1}&=0\,,\label{orbeqA1}\displaybreak[1]\\
   d\left(e^{-\phi}\ \hd{6}db-e^{\phi/2}C_1\wedge \hd{6} G_2\right)
      +d\bar{C}_3\wedge dA_1+4\kappa^2\frac{\delta \mathcal{L}_b}{\delta b}&=0\,,\label{orbeqb}\\
   d\hd{}d\phi+\frac{3}{4}e^{3\phi/2}dC_1\wedge\hd{}dC_1
      +\frac{1}{4}e^{\phi/2}d\bar{C}_3\wedge\hd{}d\bar{C}_3\qquad\qquad\qquad\nonumber&\\
      +\frac{1}{4}\Big[e^{-\phi}\ db\wedge\hd{6}db
      +\frac{1}{2}
       e^{\phi/2}\  G_2 \wedge \hd{6} G_2
      \Big]\wedge\Omega_4+2\kappa^2\frac{\delta \mathcal{L}_b}{\delta\phi}&=0\label{orbeqphi}\,,   
\end{align}
\end{subequations}
where we have defined
$\Omega_4=\delta(x^6)\cdots\delta(x^9)\ dx^6\wedge\cdots\wedge
dx^9\,$. By varying the action with respect to the metric one gets also 
 the Einstein equations that it is convenient to split into three
separate equations (according to which components of the metric are
involved in each case). 
By denoting with $x^{\rho, \sigma,...} = 
\left\{x^0,...,x^5\right\}$ the coordinates on $\mathbb{R}^{1,5}$ and
by $x^{p,q,...}=\left\{x^6,...,x^9\right\}$ the orbifolded 
coordinates we find the following equations:
\begin{subequations}\label{orbeins}
\begin{align}
  R_{\rho\sigma}-\frac{1}{2}RG_{\rho\sigma}+
    2\kappa^2\frac{\delta \mathcal{L}_b}{\delta G^{\rho\sigma}}&= 
    T^{\text{u}}_{\rho\sigma}+\Omega_4T^{\text{t}}_{\rho\sigma}\, , \\ 
  R_{pq}-\frac{1}{2}RG_{pq}+
    2\kappa^2\frac{\delta \mathcal{L}_b}{\delta G^{pq}}&= 
    T^{\text{u}}_{pq}\, , \\
  R_{q\sigma}-\frac{1}{2}RG_{q\sigma}+
    2\kappa^2\frac{\delta \mathcal{L}_b}{\delta G^{q\sigma}}&= 
    T^{\text{u}}_{q\sigma}\, . 
\end{align}
\end{subequations}
The energy-momentum tensors above refer separately to those of the 
``twisted'' and ``untwisted'' fields, and are given by:
\begin{subequations}
\begin{align}
  T^{\text{u}}_{\mu\nu}&=\frac{1}{2}(\partial_{\mu}\phi\, \partial_{\nu}\phi -
       \frac{1}{2}(\partial\phi)^2 G_{\mu\nu}) +
     \frac{1}{2}e^{3\phi/2}(F_{2\, \mu A}F_{2\, \nu}^{\ \ A} - 
       \frac{1}{4}(F_2)^2G_{\mu\nu})  \\ 
     &+\frac{1}{2\cdot3!}e^{\phi/2}
       (F_{4\, \mu ABC}F_{4\, \nu}^{\ \ ABC} - 
       \frac{1}{8}(F_4)^2G_{\mu\nu})\, , \nonumber \\  
  T^{\text{t}}_{\rho\sigma}&=\frac{1}{2}\frac{\sqrt{-G_{(6)}}}{\sqrt{-G}}\left(
     \frac{1}{2}e^{-\phi}(\partial_\rho b\, \partial_\sigma b -
       \frac{1}{2}(\partial b)^2 G_{\rho\sigma}) +
     \frac{1}{2}e^{\phi/2}(G_{2\, \rho A}G_{2\, \sigma}^{\ \ A} - 
       \frac{1}{4}(G_2)^2G_{\rho\sigma})\right)\, ,
\end{align}
\end{subequations}
where, in the expression for $T^{\text{u}}_{\mu\nu}$, indices
$\mu,\nu$ run over the appropriate coordinates (according to the
equation in which they are used) and, in all cases, summed indices
($A,B,\ldots$) run over {\em all} ten dimensional coordinates. In the
expression for $T^{\text{t}}_{\rho\sigma}$, $G_{(6)}$ refers to the
determinant of the restriction of the ten dimensional metric to the
6-dimensional subspace $\mathbb{R}^{1,5}$.

As explained in section \ref{setup}, we are interested in a bound
state of $N$ fractional D2-branes and $M$ D6-branes. The world-volume
of the D2-branes extends in the directions $x^0,x^1,x^2$, while these
branes are stuck at the orbifold fixed point $x^6=x^7=x^8=x^9=0\,$.
The D6-branes extend in the directions $x^0,x^1,x^2$ as well as along
the orbifolded directions $x^6,x^7,x^8,x^9$.

For the ``untwisted'' fields we consider the following 
 standard Ansatz for a D2/D6 system:
\begin{subequations}\label{ansatz}
\begin{align}
   ds^2&=H_2^{-5/8}H_6^{-1/8}\eta_{\alpha\beta}dx^\alpha dx^\beta
       +H_2^{3/8}H_6^{7/8}\delta_{ij}dx^idx^j
       +H_2^{3/8}H_6^{-1/8}\delta_{pq}dx^pdx^q\,,\\
   e^\phi&=H_2^{1/4}H_6^{-3/4}\,,\\
   \bar{C}_3&=\left(H_2^{-1}-1\right)dx^0\wedge dx^1\wedge dx^2\,,\label{ansC3}
\end{align}
\end{subequations}
where we have divided the coordinates in three groups: 
$x^{\alpha,\beta,\ldots}=\left\{x^0,x^1,x^2\right\}$ denote the coordinates
along the world-volume of both branes,
$x^{i,j,\ldots}=\left\{x^3,x^4,x^5\right\}$ denote the ones transverse to both, while
$x^{p,q,\ldots}=\left\{x^6,x^7,x^8,x^9\right\}$ denote the (orbifolded)
coordinates along the world-volume of the D6-branes and transverse to
the D2-branes.
The function $H_2$ depends on the radial coordinate $\rho=\sqrt{(x^3)^2+\ldots+(x^9)^2}$
of the space transverse to the D2-brane, while the function
$H_6$ depends only on the radial coordinate 
of the common transverse space $r=\sqrt{\delta_{ij}x^ix^j}\,$.

In order to find a sensible Ansatz for the fields $A_1$ and $C_1$ we
need to take a more careful look at the contributions coming from the
boundary action describing the world-volume theory of the branes. For
our system, such an action is the sum of a term describing the
D2-branes and a term describing the D6-branes:
\begin{equation}
   S_b=NS_2+MS_6\,,
\end{equation}
The relevant parts of the action (as explained in
Ref.~\cite{Bertolini:2000jy}, only the linear terms contribute) are
given by (see appendix \ref{wvact}):
\begin{subequations}
\begin{multline}\label{D2act}
   S_2=\frac{T_{2}}{2\kappa}\Bigg\{-
         \int{d^{3}x\ e^{-\phi/4}
         \sqrt{-\det G_{\alpha\beta}} 
         \left(1+\frac{\tilde{b}}{2\pi^2\alpha'}\right)} \\
         +\int_{{\cal M}_3}{\left[
         \bar{C}_3\left(1+\frac{\tilde{b}}{2\pi^2\alpha'}\right)+
         \frac{A_3}{2\pi^2\alpha'}\right]}\Bigg\}\,,
\end{multline}
\begin{multline}\label{D6act}
   S_6=\frac{T_6}{\kappa}\left\{
           -\int d^{7}x\ e^{\frac{3}{4}\phi}\sqrt{-\det G_{\rho\sigma}}
           +\int_{\mathcal{M}_{7}}C_{7}\right\}\\
           +\frac{T_2}{2\kappa}\frac{1}{2(2\pi\sqrt{\alpha'})^2}\left\{
           \int d^{3}\xi \sqrt{-\det G_{\alpha\beta}}\ \tilde{b}
           -\int_{\mathcal{M}_{3}}A_{3}\right\}+\ldots\,,
\end{multline}
\end{subequations}
where the indices $\alpha,\beta,\ldots$ run along
the common world-volume $\mathcal{M}_{3}$ and $\rho,\sigma,\ldots$ 
along the whole D6-brane
world-volume $\mathcal{M}_{7}\,$. 

We notice that the previous boundary actions do not depend on the
fields $C_1$ and $A_1$. This means that eq.s~\eqref{orbeqC1}
and \eqref{orbeqA1} will not contain the contribution coming from the
boundary action:
\begin{subequations}
\begin{align} 
   d\left(e^{3\phi/2}\ \hd{} dC_1\right)
      -\frac{1}{2}e^{\phi/2}db\wedge\hd{6}G_2\wedge\Omega_4&=0\,,
\label{ansC1}\\
   d\left(e^{\phi/2}\ \hd{6} G_2\right)
      +db\wedge d\bar{C}_3&=0\,.
\end{align}
\end{subequations}
Taking into account the expression in eq.~\eqref{ansC3} for
 $\bar{C}_3\,$, we see that the second equation 
is easily satisfied by imposing: 
\begin{equation}\label{duansA1}
   e^{\phi/2}\ \hd{6} G_2=H_2^{-1} db\wedge dx^0 \wedge dx^1 \wedge dx^2\,.
\end{equation}
Eq.~\eqref{duansA1} implies that the second term of eq.~\eqref{ansC1} vanishes. 
Then, eq.~\eqref{ansC1} can be satisfied by imposing:
\begin{equation}\label{duansC1}
    e^{3\phi/2}\ \hd{} dC_1 = -d\left(H_6^{-1}\right)
        dx^0\wedge dx^1\wedge dx^2\wedge dx^6\wedge\cdots\wedge dx^9\,.
\end{equation}
Eq.s~\eqref{duansA1} and \eqref{duansC1} imply after some
manipulations the following expressions for $A_1$ and $C_1\,$:
\begin{subequations}\label{ansatz2}
\begin{align}
   dA_1&=C_1\wedge db+\frac{1}{2}\varepsilon_{ijk}H_6\partial_ib dx^j\wedge dx^k\,,\\
   dC_1&=\frac{1}{2}\varepsilon_{ijk}\partial_iH_6 dx^j\wedge dx^k\,,
 \end{align}
\end{subequations}
where $\varepsilon_{ijk}$ is such that
$\varepsilon_{345}=\varepsilon^{345}=+1\,$. 

We are now ready to find the complete solution. 
Inserting the Ans\"atze \eqref{ansatz}-\eqref{ansatz2} into the 
equations of motion \eqref{orbeqs} and computing all the relevant 
contributions coming
from the boundary action $S_b\,$, after some algebra we get:
\begin{subequations}
\begin{equation}\label{H2}
   \left(\delta^{ij}\partial_i\partial_j 
      +H_6\delta^{pq}\partial_p\partial_q\right) H_2
      +\frac{1}{2}H_6\delta^{ij}\partial_i b\partial_j b\delta(x^6)\cdots\delta(x^9)
      +\kappa T_2N\delta(x^3)\cdots\delta(x^9)=0\,,
\end{equation}
from the eq.~\eqref{orbeqC3} for $\bar{C}_3$,
\begin{equation}\label{2eq}
   H_2^{-1}\Big(H_6\delta^{ij}\partial_i \partial_j b
      +2\delta^{ij}\partial_i H_6 \partial_j b\Big)
      -\frac{\kappa T_2}{4\pi^2\alpha'}(4N-M)\delta(x^3)\cdots\delta(x^5)=0\,,   
\end{equation}
from the eq.~\eqref{orbeqb} for $b$ and
\begin{multline}\label{3eq}
   \frac{1}{4}H_2^{-1}
      \Big(\left(\delta^{ij}\partial_i\partial_j
      +H_6\delta^{pq}\partial_p\partial_q\right) H_2
      +\frac{1}{2}H_6\delta^{ij}\partial_i b\partial_j b\delta(x^6)\cdots\delta(x^9)\Big)\\
      -\frac{3}{4}H_6^{-1}\delta^{ij}\partial_i\partial_j H_6
      +\frac{\kappa T_2}{4}\left(N-\frac{6T_6}{T_2}M\right)
      \delta(x^3)\cdots\delta(x^9)=0\,,
\end{multline}
from the eq.~\eqref{orbeqphi} for $\phi$.
\end{subequations}
Plugging eq.~\eqref{H2} into eq.~\eqref{3eq} we get the following
equation for the function $H_6$:
\begin{equation}\label{1eqH6}
   \delta^{ij}\partial_i\partial_j H_6
      =-2\kappa T_6 M \delta(x^3)\cdots\delta(x^5)\,,
\end{equation}
whose solution is given by the following harmonic function:
\begin{equation}
   H_6(r)=1+\frac{g_s\sqrt{\alpha'}M}{2r}\,.
\end{equation}
Analogously, using eq.~\eqref{1eqH6} into eq.~\eqref{2eq} we obtain
the following equation for the function $Z\equiv H_6b$:
\begin{equation}
   \delta^{ij}\partial_i\partial_j Z
      =\frac{\kappa T_2}{2\pi^2\alpha'} 
      (2N-M) \delta(x^3)\cdots\delta(x^5)\,, 
\end{equation}
which is solved by:
\begin{equation}
   Z(r)=\frac{\left(2\pi\sqrt{\alpha'}\right)^2}{2}
      \left(1-\frac{g_s\sqrt{\alpha'}(2N-M)}{r}\right)\,,
\end{equation}
where we have chosen the constant term in order to satisfy the
condition \eqref{asp2} for the background value of the field $b$.
We are now left with eq.~\eqref{H2}, which is in general difficult to 
solve. 
Finally, after some computation one can show 
that with our Ansatz the equations \eqref{orbeins} 
for the metric are also satisfied provided that eq.~\eqref{H2} holds.

Finally, the fields $C_1$ and $A_1$ are obtained by integrating 
eq.s~(\ref{ansatz2}). In order to do so, we change the coordinate system
into polar coordinates in the common transverse space: 
$\left(x^3,x^4,x^5\right)\longrightarrow\left(r,\theta,\varphi\right)$.
Then we obtain:
\begin{subequations}
\begin{align}
   C_1&=\frac{g_s\sqrt{\alpha'}M}{2}\cos\theta d\varphi\,,\\
   A_1&=-\pi^2\alpha'\frac{g_s\sqrt{\alpha'}(4N-M)}
      {1+\frac{g_s\sqrt{\alpha'}M}{2r}}  \cos\theta d\varphi\,.
\end{align}
\end{subequations}
The supergravity solution that we have found is summarized in 
eq.~\eqref{solution}.


\section{World-volume actions for fractional branes}\label{wvact}

The world-volume action for a fractional D$p$-brane ($p\leq5$) transverse to the
orbifold space $\mathbb{R}^4/\mathbb{Z}_2$ can be obtained in several ways. 
Recalling that a fractional D$p$-brane is a D$(p+2)$-brane 
wrapped\footnote{To
be precise, we are considering fractional branes of ``type 1'', which have a $B$-flux
on the shrinking cycle but not an $F$-flux.} on the vanishing cycle
$\Sigma_2$ defined in section \ref{setup}, one can get the action for
a fractional D$p$-brane 
starting from the one of a D$(p+2)$-brane, which in the Einstein frame is:
\begin{equation}
   S_{p+2}=S_{\text{DBI}}+S_{\text{WZ}}\,,
\end{equation}
with:
\begin{subequations}\label{DBIWZ}
\begin{align}
   S_{\text{DBI}}&=-\frac{T_{p+2}}{\kappa}
         \int d^{p+3}\xi e^{\frac{p-1}{4}\phi}
         \sqrt{-\det \big[G_{ab}+e^{-\frac{\phi}{2}}
         \left(B_{ab}+2\pi\alpha'F_{ab}\right)\big]}\,,\label{DBI}\\
   S_{\text{WZ}}&=\frac{T_{p+2}}{\kappa}
         \int_{\mathcal{M}_{p+3}}
         \sum_q C_q \wedge e^{B+2\pi\alpha'F}\,,
\end{align}
\end{subequations}
where $\xi^{a,b,\ldots}=\{\xi^0,\ldots,\xi^{p+2}\}$ 
are the coordinates of the brane world-volume and
$T_p=\sqrt{\pi}(2\pi\sqrt{\alpha'})^{3-p}\,$. All bulk fields
in eq.s~\eqref{DBIWZ} are pullbacks onto the world-volume
$\mathcal{M}_{p+3}$ of the brane.

Let us start by considering the DBI part of the action. 
In order to wrap the brane on the cycle $\Sigma_2$ we have to
impose the decomposition in eq.~\eqref{decomp} for the field $B_2$ (we 
suppose
that it has no components outside the cycle). The metric has no
support on $\Sigma_2$, so eq.~\eqref{DBI} becomes:
\begin{equation}
\begin{split}
   S_{\text{DBI}}&=-\frac{T_{p+2}}{\kappa}
         \int d^{p+1}\xi e^{\frac{p-1}{4}\phi}
         \sqrt{-\det \big[G_{\alpha\beta}+e^{-\frac{\phi}{2}}
         2\pi\alpha'F_{\alpha\beta}\big]}
         \ e^{-\frac{\phi}{2}}
         \int_{\Sigma_2}b\omega_2\,,\\
         &=-\frac{T_{p}}{2\kappa}
         \int d^{p+1}\xi e^{\frac{p-3}{4}\phi}
         \sqrt{-\det \big[G_{\alpha\beta}+e^{-\frac{\phi}{2}}
         2\pi\alpha'F_{\alpha\beta}\big]}
         \left(1+\frac{\tilde{b}}{2\pi^2\alpha'}\right)\,,
\end{split}
\end{equation}
where we have used eq.s~(\ref{asp}-\ref{asp2}) and the relation
$T_{p}=(2 \pi \sqrt{\alpha'})^2 T_{p+2}\,$. The coordinates
$\xi^{\alpha,\beta,\ldots}=\{\xi^0,\ldots,\xi^{p}\}$ are the
coordinates of the world-volume $\mathcal{M}_p$ of the fractional
D$p$-brane. Turning to the WZ part, we have to decompose the R-R
potentials in a similar fashion as in eq.~\eqref{decomp}:
\begin{equation}
   C_q=\bar{C}_q+A_{q-2}\wedge\omega_2
\end{equation}
(notice that $\bar{C}_{p+3}$ vanishes). To discuss what is
the result of wrapping,
let us first consider the highest rank field $C_{p+3}$, whose contribution
to the WZ action is given by:
\begin{equation}
   \frac{T_{p+2}}{\kappa}\int_{\mathcal{M}_{p+3}}C_{p+3}
      =\frac{T_{p+2}}{\kappa}\int_{\mathcal{M}_{p+3}}A_{p+1}\wedge\omega_2
      =\frac{T_{p}}{2\kappa}\int_{\mathcal{M}_{p+1}}\frac{A_{p+1}}{2\pi^2\alpha'}\,,
\end{equation}
where we have used eq.s~\eqref{omegaprop}. Considering 
now $C_{p+1}$, one gets:
\begin{equation}
   \frac{T_{p+2}}{\kappa}\int_{\mathcal{M}_{p+3}}C_{p+1}\wedge B
      \longrightarrow\frac{T_{p+2}}{\kappa}\int_{\mathcal{M}_{p+3}}\bar{C}_{p+1} b\wedge\omega_2
      =\frac{T_{p}}{2\kappa}\int_{\mathcal{M}_{p+1}}
      \bar{C}_{p+1}\left(1+\frac{\tilde{b}}{2\pi^2\alpha'}\right)\,.
\end{equation}
Therefore the first term of the WZ action is:
\begin{equation}
   \frac{T_{p}}{2\kappa}\int_{\mathcal{M}_{p+1}}\left[
      \bar{C}_{p+1}\left(1+\frac{\tilde{b}}{2\pi^2\alpha'}\right)
      +\frac{A_{p+1}}{2\pi^2\alpha'}\right]\,.
\end{equation}
If we consider any other lower rank potential, one can see that the 
relevant
contributions to the WZ action always involve the following combinations
of fields:
\begin{equation}
   \mathcal{C}_q=
     \bar{C}_q\left(1+\frac{\tilde{b}}{2\pi^2\alpha'}\right)
      +\frac{A_q}{2\pi^2\alpha'}\,,
\end{equation}
This means that the world-volume action for a fractional 
D$p$-brane can always be put in the form:
\begin{equation}\label{FracDp}
   S_{p}=S_{\text{DBI}}+S_{\text{WZ}}\,,
\end{equation}
where:
\begin{subequations}
\begin{align}
   S_{\text{DBI}}&=-\frac{T_{p}}{2\kappa}
         \int d^{p+1}\xi e^{\frac{p-3}{4}\phi}
         \sqrt{-\det \big[G_{\alpha\beta}+e^{-\frac{\phi}{2}}
         2\pi\alpha'F_{\alpha\beta}\big]}
         \left(1+\frac{\tilde{b}}{2\pi^2\alpha'}\right)\,,\\
   S_{\text{WZ}}&=\frac{T_{p}}{2\kappa}
         \int_{\mathcal{M}_{p+1}}
         \sum_q \mathcal{C}_q \wedge e^{2\pi\alpha'F}\,.
\end{align}
\end{subequations}
The precise expression of the action for a fractional D$p$-brane is 
confirmed by the
couplings of the brane to the bulk fields, computed with the boundary
state formalism \cite{Bertolini:2001dk} and with explicit computation
of string scattering amplitudes on a disk \cite{Merlatti:2001ne}.

In this paper we also consider D-branes whose
world-volume directions extend along the whole orbifold space, namely
D$(p+4)$-branes with four longitudinal directions along $x^6,\ldots,x^9$ 
and $p+1$ along $x^0,\ldots,x^p\,$. In this case the terms linear in
the bulk fields of the
boundary action can be inferred from the couplings computed with
the boundary state\footnote{One
has also to take into account the fact that the
boundary state sees the fields correctly normalized on
the covering space, while we are using fields that are
correctly normalized on the orbifold \cite{Billo:2001vg}.} 
\cite{Bertolini:2001qa}, and one gets:
\begin{multline}\label{Dp+4} 
   S_{p+4}=\frac{T_{p+4}}{\kappa}\left\{
           -\int d^{p+5}\xi e^{\frac{p+1}{4}\phi}\sqrt{-\det G_{\rho\sigma}}
           +\int_{\mathcal{M}_{p+5}}C_{p+5}\right\}\\
           +\frac{T_{p}}{2\kappa}\frac{1}{2(2\pi\sqrt{\alpha'})^2}\left\{
           \int d^{p+1}\xi \sqrt{-\det G_{\alpha\beta}}\ \tilde{b}
           -\int_{\mathcal{M}_{p+1}}A_{p+1}\right\}+\ldots\,,
\end{multline}
where $x^{\rho,\sigma,\ldots}$ are the coordinates of the brane
world-volume $\mathcal{M}_{p+5}\,$, while
$x^{\alpha,\beta,\ldots}$ are the coordinates along the part
$\mathcal{M}_{p+1}$ of
the world-volume which lie outside the orbifold directions. The ellipses
in the action \eqref{Dp+4} stand for terms of higher order in the fields,
not accounted by the boundary state approach.

\section{The running coupling constant of $\mathcal{N}=4\,$, $D=2+1$
 SYM theory}\label{gaugerun}

In this appendix, we briefly compute the expression of the
running gauge coupling constant of $\mathcal{N}=4\,$, $D=2+1$
super Yang--Mills theory, in order to compare the perturbative
gauge theory result \cite{Fradkin:1983kf}
(see Ref.~\cite{DiVecchia:1998ky} for a review) with what we obtain 
from the supergravity solutions
for the wrapped and the fractional brane systems.

The one-loop effective action for a $D$-dimensional field theory 
expanded around a background
which is a solution of the classical field equations can be expressed
as\footnote{A bar on an operator in eq.s~\eqref{effact} and
\eqref{deter} indicates that the operator is evaluated at the
background value $\bar{A}_\mu^a$ of the gauge field.}:
\begin{equation}\label{effact}
   S_{\text{eff}}=\frac{1}{4g_{\text{YM}}^2}\int d^Dx\left\{
      \bar{F}^a_{\mu\nu}\bar{F}^a_{\mu\nu}+\frac{1}{2}\tr\log\Delta_1
      +\left(\frac{N_s}{2}-1\right)\tr\log\Delta_0
      -N_f\tr\log\Delta_{1/2}\right\}\,, 
\end{equation}
where $N_s$ and $N_f$ are respectively the number of scalars and
Dirac fermions and where:
\begin{equation}\label{deter}
   \left(\Delta_1\right)^{ab}_{\mu\nu}=-\left(\bar{D}^2\right)^{ab}\delta_{\mu\nu}
      +2f^{acb}\bar{F}^c_{\mu\nu}\,,\quad
   \left(\Delta_0\right)^{ab}=-\left(\bar{D}^2\right)^{ab}\,,\quad
   \Delta_{1/2}=i\bar{\,/\hspace{-0.29cm}D}\,,
\end{equation}
$D_{\mu}$ being the covariant derivative and $f^{abc}$
the gauge group structure constants.
The part of the determinants in eq.~\eqref{effact} quadratic in the gauge
fields can be extracted obtaining:
\begin{equation}\label{effrun}
   S_{\text{eff}}=\frac{1}{4}\int d^Dx F^2\left\{\frac{1}{g^2_{\text{YM}}}+I\right\}\,,
\end{equation}
where:
\begin{equation}
   I=\frac{1}{(4\pi)^{D/2}}\int_0^\infty \frac{ds}{s^{D/2-1}}e^{-\mu^2s}R\,,
\end{equation}
where $\mu$ is the mass of the fields, and
\begin{equation}
   R=2\left[\frac{N_s}{12}c_s+\frac{D-26}{12}c_v+\frac{2^{[D/2]}N_f}{6}c_f\right]\,,
\end{equation}
where $[D/2]=D/2$ if $D$ is even and $[D/2]=\tfrac{D-1}{2}$ if $D$ is
odd, and where the constants $c$ set the normalization of the
generators of the gauge group ($Tr( \lambda^a \lambda^b ) = c \delta^{ab}$)
in the representations under which the
scalars, the vector and the fermions respectively
transform. Concentrating on the case $D=3$, we get:
\begin{equation}\label{I3d}
   I=\frac{1}{(4\pi)^{3/2}}\int_0^\infty\frac{ds}{s^{1/2}}e^{-\mu^2s}R
      =\frac{1}{8\pi\mu}R\,,
\end{equation}
with
\begin{equation}\label{R3d}
   R=2\left[\frac{N_s}{12}c_s-\frac{23}{12}c_v+\frac{N_f}{3}c_f\right]\,.
\end{equation}
In our case we have a theory with gauge group $SU(N)$
and 8 supercharges,
coupled to $M$ hypermultiplets in its fundamental
representation.
The vector multiplet contains 2 Dirac fermions and
3 scalars, while each hypermultiplet is made up of 2 Dirac fermions and
4 scalars.
Recalling that $c=\tfrac{1}{2}$ for the fundamental representation
and $c=N$ for the adjoint representation, eq.~\eqref{R3d} gives:
\begin{equation}
   R=-2N+M\,,
\end{equation}
and the running effective coupling given by 
eq.s~\eqref{effrun}-\eqref{I3d} is equal to:
\begin{equation}
      \frac{1}{g^2_{\text{YM}}(\mu)}=
   \frac{1}{g^2_{\text{YM}}}
   \left(1-g^2_{\text{YM}}\frac{2N-M}{8\pi\mu}\right)\,,
\end{equation}
This expression of the one-loop running coupling constant is in
complete agreement 
with both our results in eq.s~\eqref{wrunning} and \eqref{running}.




\begin{thebibliography}{50}

\bibitem{Maldacena:1997re}
J.~Maldacena,
``The large $N$ limit of superconformal field theories and supergravity'',
Adv.\ Theor.\ Math.\ Phys.\  {\bf 2} (1998) 231
[Int.\ J.\ Theor.\ Phys.\  {\bf 38} (1998) 1113],
hep-th/9711200.

\bibitem{MN}
J. Maldacena and C. N\'u\~nez,  
``Supergravity description of field theories on curved ma\-nifolds and a 
no-go theorem'',
Int.\ J.\ Mod.\ Phys. {\bf A16} (2001) 82, 
hep-th/0007018.

\bibitem{Douglas:1996sw}
M.~R.~Douglas and G.~W.~Moore,
``D-branes, Quivers, and ALE Instantons'',
hep-th/9603167.

\bibitem{Johnson:1996py}
C.~V.~Johnson and R.~C.~Myers,
``Aspects of type IIB theory on ALE spaces'',
Phys.\ Rev.\ D {\bf 55} (1997) 6382,
hep-th/9610140.

\bibitem{Diaconescu:1998br}
D.~Diaconescu, M.~R.~Douglas and J.~Gomis,
``Fractional branes and wrapped branes'',
JHEP {\bf 9802} (1998) 013,
hep-th/9712230.

\bibitem{Klebanov:1999rd}
I.~R.~Klebanov and N.~A.~Nekrasov,
``Gravity duals of fractional branes and logarithmic RG flow'',
Nucl.\ Phys.\ B {\bf 574} (2000) 263,
hep-th/9911096.

\bibitem{Herzog:2001xk}
C.~P.~Herzog, I.~R.~Klebanov and P.~Ouyang,
``Remarks on the warped deformed conifold'',
hep-th/0108101.

\bibitem{Fayyazuddin:1999zu}
A.~Fayyazuddin and D.~J.~Smith,
``Localized intersections of M5-branes and four-dimensional  superconformal field theories'',
JHEP {\bf 9904} (1999) 030,
hep-th/9902210.

\bibitem{Fayyazuddin:2000em}
A.~Fayyazuddin and D.~J.~Smith,
``Warped AdS near-horizon geometry of completely localized intersections  of M5-branes'',
JHEP {\bf 0010} (2000) 023,
hep-th/0006060.

\bibitem{Brinne:2000fh}
B.~Brinne, A.~Fayyazuddin, S.~Mukhopadhyay and D.~J.~Smith,
``Supergravity M5-branes wrapped on Riemann surfaces and their QFT duals'',
JHEP {\bf 0012} (2000) 013,
hep-th/0009047.

\bibitem{Bertolini:2001dk}
M.~Bertolini, P.~Di Vecchia, M.~Frau, A.~Lerda, R.~Marotta and I.~Pesando,
``Fractional D-branes and their gauge duals'',
JHEP {\bf 0102} (2001) 014,
hep-th/0011077.

\bibitem{Polchinski:2001mx}
J.~Polchinski,
``N = 2 gauge-gravity duals'',
Int.\ J.\ Mod.\ Phys.\ A {\bf 16} (2001) 707,
hep-th/0011193.

\bibitem{Billo:2001vg}
M.~Bill\`o, L.~Gallot and A.~Liccardo,
``Classical geometry and gauge duals for fractional branes on ALE  orbifolds'',
hep-th/0105258.

\bibitem{Grana:2001xn}
M.~Gra\~na and J.~Polchinski,
``Gauge / gravity duals with holomorphic dilaton'',
hep-th/0106014.

\bibitem{Bertolini:2001qa}
M.~Bertolini, P.~Di Vecchia, M.~Frau, A.~Lerda and R.~Marotta,
``N = 2 gauge theories on systems of fractional D3/D7 branes'',
hep-th/0107057.

\bibitem{Gauntlett:2001ps}
J.~P.~Gauntlett, N.~Kim, D.~Martelli and D.~Waldram,
``Wrapped fivebranes and N = 2 super Yang--Mills theory'',
Phys.\ Rev.\ D {\bf 64} (2001) 106008,
hep-th/0106117.

\bibitem{Bigazzi:2001aj}
F.~Bigazzi, A.~L.~Cotrone and A.~Zaffaroni,
``N = 2 gauge theories from wrapped five-branes'',
Phys.\ Lett.\ B {\bf 519} (2001) 269,
hep-th/0106160.

\bibitem{Maldacena:2000yy}
J.~M.~Maldacena and C.~N\'u\~nez,
``Towards the large N limit of pure N = 1 super Yang Mills'',
Phys.\ Rev.\ Lett.\  {\bf 86} (2001) 588,
hep-th/0008001.

\bibitem{Nunez:2001pt}
C.~N\'u\~nez, I.~Y.~Park, M.~Schvellinger and T.~A.~Tran,
``Supergravity duals of gauge theories from F(4) gauged supergravity in  six dimensions'',
JHEP {\bf 0104} (2001) 025,
hep-th/0103080.

\bibitem{Gomis:2001vk}
J.~Gomis,
``D-branes, holonomy and M-theory'',
Nucl.\ Phys.\ B {\bf 606} (2001) 3,
hep-th/0103115.

\bibitem{Edelstein:2001pu}
J.~D.~Edelstein and C.~N\'u\~nez,
``D6 branes and M-theory geometrical transitions from gauged  supergravity'',
JHEP {\bf 0104} (2001) 028,
hep-th/0103167.

\bibitem{Gomis:2001vg}
J.~Gomis and T.~Mateos,
``D6 branes wrapping K\"ahler four-cycles'',
hep-th/0108080.

\bibitem{Gomis:2001aa}
J.~Gomis and J.~G.~Russo,
``D = 2+1 N = 2 Yang--Mills theory from wrapped branes'',
hep-th/0109177.

\bibitem{Gauntlett:2001ur}
J.~P.~Gauntlett, N.~Kim, D.~Martelli and D.~Waldram,
``Fivebranes wrapped on SLAG three-cycles and related geometry'',
hep-th/0110034.

\bibitem{Johnson:2000qt}
C.~V.~Johnson, A.~W.~Peet and J.~Polchinski,
``Gauge theory and the excision of repulson singularities'',
Phys.\ Rev.\ D {\bf 61} (2000) 086001,
hep-th/9911161.

\bibitem{Wijnholt:2001us}
M.~Wijnholt and S.~Zhukov,
``Inside an enhan\c con: Monopoles and dual Yang--Mills theory'',
hep-th/0110109.

\bibitem{Petrini:2001fk}
M.~Petrini, R.~Russo and A.~Zaffaroni,
``N = 2 gauge theories and systems with fractional branes'',
Nucl.\ Phys.\ B {\bf 608} (2001) 145,
hep-th/0104026.

\bibitem{Johnson:2001wm}
C.~V.~Johnson, R.~C.~Myers, A.~W.~Peet and S.~F.~Ross,
``The enhan\c con and the consistency of excision'',
Phys.\ Rev.\ D {\bf 64} (2001) 106001,
hep-th/0105077.

\bibitem{Merlatti:2001gd}
P.~Merlatti,
``The enhan\c con mechanism for fractional branes'',
hep-th/0108016.

\bibitem{Alvarez-Gaume:1981hm}
L.~Alvarez-Gaum\'e and D.~Z.~Freedman,
``Geometrical Structure And Ultraviolet Finiteness In The Supersymmetric Sigma Model'',
Commun.\ Math.\ Phys.\  {\bf 80} (1981) 443.

\bibitem{Seiberg:1996nz}
N.~Seiberg and E.~Witten,
``Gauge dynamics and compactification to three dimensions'',
hep-th/9607163.

\bibitem{BVS}
M.~Bershadsky, C.~Vafa and V.~Sadov, 
``D-Branes and Topological Field Theories'', 
Nucl.\ Phys.\ {\bf B463} (1996) 420, 
hep-th/9511222.

\bibitem{Cv1}
M.~Cveti\v c, M.~J.~Duff, P.~Hoxha, J.~T.~Liu, H.~L\"u, J.~X.~Lu, 
R.~Mart\'{\i}nez-Acosta, C.~N.~Pope, H.~Sati and T.~A.~Tran, 
``Embedding of AdS Black Holes in Ten and Eleven Dimensions'',
Nucl.\ Phys.\ {\bf B558} (1999) 96, 
hep-th/9903214.

\bibitem{Cv2}
M.~Cveti\v c, H. L\"u and C.~N.~Pope, 
``Consistent Kaluza-Klein Sphere Reductions'',
Phys.\ Rev. {\bf D62} (2000) 064028,
hep-th/0003286.

\bibitem{Gauntlett:2001qs}
J.~P.~Gauntlett, N.~Kim, S.~Pakis and D.~Waldram,
``Membranes wrapped on holomorphic curves'',
hep-th/0105250.

\bibitem{Babington:2001nh}
J.~Babington and N.~Evans,
``Field Theory Operator Encoding in N=2 Geometries'',
hep-th/0111082.

\bibitem{Eguchi:1978gw}
T.~Eguchi and A.~J.~Hanson,
``Selfdual Solutions To Euclidean Gravity'',
Annals Phys.\  {\bf 120} (1979) 82.

\bibitem{Johnson:2000ch}
C.~V.~Johnson,
``D-brane primer'',
hep-th/0007170.

\bibitem{Hawking:1977jb}
S.~W.~Hawking,
``Gravitational Instantons'',
Phys.\ Lett.\ A {\bf 60} (1977) 81.

\bibitem{Frau:2001gk}
M.~Frau, A.~Liccardo and R.~Musto,
``The geometry of fractional branes'',
Nucl.\ Phys.\ B {\bf 602} (2001) 39,
hep-th/0012035.

\bibitem{Aspinwall:1995zi}
P.~S.~Aspinwall,
``Enhanced gauge symmetries and K3 surfaces'',
Phys.\ Lett.\ B {\bf 357} (1995) 329,
hep-th/9507012.

\bibitem{Douglas:1996xg}
M.~R.~Douglas,
``Enhanced gauge symmetry in M(atrix) theory'',
JHEP {\bf 9707} (1997) 004,
hep-th/9612126.

\bibitem{Cvetic:2000mh}
M.~Cveti\v c, H.~L\"u and C.~N.~Pope,
``Brane resolution through transgression'',
Nucl.\ Phys.\ B {\bf 600} (2001) 103,
hep-th/0011023.

\bibitem{Schmidhuber:1996}
C. Schmidhuber, 
``D-brane actions'',
Nucl.\ Phys.\  {\bf B467} (1996) 146,
hep-th/9601003.

\bibitem{Bertolini:2000jy}
M.~Bertolini, P.~Di Vecchia, M.~Frau, A.~Lerda, R.~Marotta and R.~Russo,
``Is a classical description of stable non-BPS D-branes possible?'',
Nucl.\ Phys.\ B {\bf 590} (2000) 471,
hep-th/0007097.

\bibitem{Merlatti:2001ne}
P.~Merlatti and G.~Sabella,
``World volume action for fractional branes'',
Nucl.\ Phys.\ B {\bf 602} (2001) 453,
hep-th/0012193.

\bibitem{Fradkin:1983kf}
E.~S.~Fradkin and A.~A.~Tseytlin,
``Quantum Properties Of Higher Dimensional And Dimensionally Reduced Supersymmetric Theories'',
Nucl.\ Phys.\ B {\bf 227} (1983) 252.

\bibitem{DiVecchia:1998ky}
P.~Di Vecchia,
``Duality in N = 2,4 supersymmetric gauge theories'',
hep-th/9803026.

\end{thebibliography}
\end{document}